\documentclass[]{interact}

\usepackage{epstopdf}
\usepackage{subfigure}

\usepackage{natbib}
\bibpunct[, ]{(}{)}{;}{a}{}{,}

\theoremstyle{plain}

\theoremstyle{definition}

\theoremstyle{remark}

\usepackage{color}
\usepackage{fancyvrb}

\DefineVerbatimEnvironment{Highlighting}{Verbatim}{commandchars=\\\{\}}
\usepackage{framed}
\definecolor{shadecolor}{RGB}{248,248,248}

\usepackage{hyperref}
\usepackage[utf8]{inputenc}

\usepackage{amsfonts}
\usepackage{graphicx,xfrac}
\usepackage{amsgen,amsmath,amstext,amsbsy,amsopn,amssymb,bbm}

\usepackage{setspace}
\newcommand{\X}{\mathbf{X}}

\newcommand{\Bbeta}{\boldsymbol{\beta}}

\begin{document}

\title{Spatial Analysis of the Association between School Proximity and Crime in Philadelphia}

\author{\name{Leonardo de Castro Harth$^{a}$, Bangxi Xiao$^{a}$, Shane T. Jensen$^{b}$}
\affil{$^{a}$Wharton Social Impact Initiative, The Wharton School, University of Pennsylvania, Philadelphia, PA, USA; $^{b}$Department of Statistics, The Wharton School, University of Pennsylvania, Philadelphia, PA, USA}
}

\thanks{Corresponding Author: S.T. Jensen, email: \href{mailto:stjensen@wharton.upenn.edu}{\nolinkurl{stjensen@wharton.upenn.edu}}}


\maketitle

\begin{abstract}
We use high resolution data to investigate the association between crime incidence and proximity to different types of public schools over the past fifteen years in the city of Philadelphia.  We employ two statistical methods, regression modeling and propensity score matching, in order to better isolate the association between crime and school proximity while controlling for the demographic, economic, land use and disorder characteristics of the surrounding neighborhood.  With both of these approaches, we find significantly increased crime incidence near to public schools regardless of crime outcome, educational level and time period.  The effect of school proximity on crime varies substantially depending on whether or not school is in session, as well as between different types of crime and educational levels of the school.    We see the largest effects of school proximity on crime for violent crimes near to high schools during their in-session time periods. Our results support several theories which suggest that crime should be elevated near to schools, as well as finding significant associations between crime and other aspects of the surrounding area.  \end{abstract}


\section{Introduction}

The recent availability of data gives us the opportunity to investigate characteristics of the urban environment at a higher level of resolution than previously possible.  In this paper, we focus on the extent to which proximity to different types of public schools is associated with elevated or reduced levels of crime in the surrounding neighborhood, both within and outside of time periods where schools are in session.  



There are many theories in urban planning and criminology that provide mechanisms for how schools can affect criminal incidence in their surrounding neighborhoods.   In their routine activities theory, \cite{CohFel79} state that "criminal acts require convergence in space and time of likely offenders, suitable targets and the absence of capable guardians against crime” (p.88).  Schools collect large numbers of young individuals that could serve as either perpetrators or victims in particular locations within a neighborhood for particular periods of each week day.  Within schools themselves, there is a high level of guardianship but elevated crime incidence might be expected in the surrounding neighborhood during travel to/from school when there could also be a gap in guardianship since school hours tend to end prior to work hours.   The concentration of potential perpetrators and victims together would also be predicted to elevate crime incidence under the criminal opportunity theory of \cite{Coo86}.  

Schools also gather together individuals at the age where they have the highest propensity to be either victims or perpetrators of crime.  Rates of both criminal offending and victimization are highest during teenage years, with the peak occurring during ages 16 to 19 \citep{Far86}.   This observation suggests that crime incidence would be most elevated around high schools compared to other levels of public schools.  Social interactions theory \citep{GlaSacSch96} suggests that schools increase the potential for crime through the influence of offenders on their peers. 

Many of these previous theories are collected into the larger framework of {\it pattern theory} \citep{BraBra93}, which considers criminal behavior as a multi-faceted decision process where variation in crime patterns is driven by both individual and environmental factors that vary in both space and time.  Specifically, criminal activity is impacted by the nature of the crime itself, the readiness and willingness of individuals to commit a crime, and the structural and activity backcloth of the place and time.   

Schools collect young people that commute to/from and congregate nearby the school at regular times during each week, during which time those individuals gain knowledge about the level of guardianship in the surrounding neighborhood.  In the terminology of \cite{BraBra93}, the structural and activity backcloth of schools create a template that allows for both planned criminal activities (since the surrounding neighborhood becomes well known) and spontaneous criminal activities if there is a gap in guardianship.  However, we should note that this activity backcloth is only present around school hours on weekdays and so elevated crime incidence would not be expected on the weekend or during summer months.    

Taken together, these theories suggest the following hypotheses that we will evaluate using all public schools in Philadelphia.  First, we hypothesize that neighborhoods with public schools will have elevated crime incidence relative to neighborhoods without public schools but only in time periods around and during school sessions.   Second, high schools will have more elevated crime incidence compared to middle or elementary schools, since previous work suggests crime rates peak in the late teenage years \citep{Far86}.  

Finally, we also will make a distinction in our analyses between different types of crime (violent vs. property vs. drug) as criminal incidence is impacted by the nature of the crime itself \citep{BraBra93}.  However, theory and previous empirical work (which we review below) is ambiguous on which types of crime are expected to be most impacted by proximity to a public school.  

There have also been many previous empirical investigations of crime incidence in areas near to schools.   \cite{RonLob83} find higher incidence of burglaries and auto thefts around public (but not private) high schools in 1970 in San Diego, California. In a spatial study of Omaha, Nebraska in 2000-2002, \cite{MurSwa13} find higher incidence of auto thefts and aggravated assaults in blocks containing public high schools as well as higher incidence of aggravated assaults in blocks containing public middle schools.  However, there were several other crime types examined that showed no relationship with school proximity.  Both of these studies use US census blocks as the geographical unit for their analyses and we will also focus on US census blocks in our investigation.   

Several studies have examined the impact of schools being in session on crime.  \cite{JacLef03} examine 29 US cities and find that violent crime increases while property crime decreases on days when school is in session.  \cite{BilPhi17} and \cite{BilDemRos19} find that concentrating economically disadvantaged students into the same schools in the Charlotte region was associated with increased crime, which provides empirical support for the social interation theory of \citep{GlaSacSch96}. 

There have also been several studies into changes in crime after either the opening of new schools or closing of existing schools. \cite{BriGar09} found increases in crime after the closing of neighborhood Catholic schools in Chicago whereas \cite{MacNicUke18} found no evidence of a change in crime around school openings in Philadelphia and that that changes in the number of schools are not associated with changes in crime at the census tract level.  In comparison to \cite{MacNicUke18}, we will explore associations between crime and school proximity in Philadelphia at the census block level which is a much higher resolution than census tracts, but we will not take into account the age of those public schools or opening of new public schools within the time frame of our data.

\cite{SteUkeMac19} used closures of public schools in Philadelphia to identify differences in crime rates and found significant declines in crime following school closures, which is in contrast with the lack of change around newly opened schools found by \cite{MacNicUke18}. In Philadelphia, and more generally, schools that are closed tend to be among the worst performing schools in the city, and so the study of \cite{SteUkeMac19} is focused on academically underperforming schools.   In comparison to \cite{SteUkeMac19}, our investigation focuses on all public schools in Philadelphia rather than just academically underperforming schools that were recently closed.  Thus, our study is based on a larger number of 325 public schools (vs. less than 40 closed schools) and our evaluation of crime incidence will be based on schools across the entire range of academic performance in Philadelphia.   We will revisit the findings of \cite{MacNicUke18} and \cite{SteUkeMac19} in comparison with our own results in Section~\ref{discussion}.  


Beyond schools, \cite{Mac15} and \cite{Macbrasto19} reviews previous research on different aspects of the urban environment and safety, where many quasi-experimental studies have shown that changes in green space, housing, zoning and public transit have an association with crime. Historically, it has been difficult to alter the urban environment in a well controlled experimental way and this is particularly true for public school locations.  
In an ideal controlled experiment, we could use randomization to create a balanced comparison between neighborhoods with public schools and neighborhoods without public schools.  However, public school locations are not randomly chosen.  Rather, public school locations are determined by historical convenience, transit options, land availability and economic factors as well as a myriad of other considerations.  Due to these considerations, the neighborhoods with public schools are likely to have quite different characteristics compared to neighborhoods without public schools.

Any differences in surrounding neighborhood characteristics are a challenge to our efforts to isolate the association between public schools and crime incidence.   We will be comparing crime levels between neighborhoods that are in close proximity to public schools (our ``treatment" group) and neighborhoods that are not in close proximity to public schools (our ``control" group).  However, when doing these comparisons, we must be careful to take into account the systematic differences between these two groups in terms of their surrounding neighborhood characteristics.  

In order to account for surrounding neighborhood context, we incorporate available data on the economic, demographic and land use characteristics of Philadelphia neighborhoods into our analyses.  We also incorporate additional ``Broken Windows" features \citep{WilKel03} of these neighborhoods based on available litter, code violation and 311 request data from the City of Philadelphia.  In Section~\ref{eda}, we will see substantial differences on many of these neighborhood characteristics between areas around public schools versus areas without public schools.  

We will employ two statistical techniques, regression modeling and propensity score matching, to estimate the association between crime and public school locations while controlling for surrounding demographic, economic and land use measures.   We will fit a set of regression models to estimate the partial effects of proximity to different levels of public schools on crime while controlling for other neighborhood factors by including them in our regression models.  


In an alternative approach, we also address the systematic differences in surrounding neighborhood context by a careful matching of individual neighborhoods that contain a public school to individual neighborhoods that do not contain a public school but have highly similar surrounding neighborhood context.  Specifically, we will use propensity score matching \citep{RosRub83} to create matched pairs of neighborhoods where each pair of consists of a neighborhood with a public school and a neighborhood without a public school.  The matching is based on our collected demographic, economic, land use, and disorder measures so that our within-pair comparisons of crime are balanced in terms of all these aspects of surrounding neighborhood context.

With both statistical approaches,  we find significantly increased crime incidence near to public schools regardless of crime outcome, educational level and time period.  However, our estimated school effects vary substantially between different crime outcomes, educational levels and time periods. Other estimated parameters in our regression model support both theory \citep{WilKel03} and previous empirical studies \citep{BerBlo09} of the association between crime and characteristics of the urban environment.  

We describe our available data on public school locations, crime, demographic, economic characteristics, land use, litter, code violations and 311 requests for Philadelphia in Section~\ref{data}, along with some exploratory analysis of this data in Section~\ref{eda}.  In Section~\ref{regression}, we fit a series of regression models to estimate the association between crime, proximity to public schools and other neighborhood factors.  Our propensity score matching analysis in Section~\ref{matching} uses matched pairs of neighborhoods to evaluate the association between crime and proximity to different levels of public schools.  We conclude with a brief summary and discussion in Section~\ref{discussion}.  

\section{Urban Data in Philadelphia}\label{data}

Our analyses will be based on publicly available data on school locations, crime, and the economic, demographic land use and disorder characteristics of each Philadelphia neighborhood.  We also incorporate a comprehensive database on business types and locations that has been compiled and curated by our research group.  Below we provide individual descriptions of each data source and the quantitative measures we create for local neighborhoods in Philadelphia.    

Data processing and analyses were performed in the R statistical computing environment \citep{R21}.  The github repository for our code and data can be accessed at: 
\begin{center}
\url{https://github.com/leonardoharth/School_Crime_Research}
\end{center}

\subsection{School Data}\label{schooldata}

We have the GPS coordinates for the location of every public school in the city of Philadelphia.  We also have the educational level for each school, e.g.  elementary vs. middle vs. high school.  Some public schools contain multiple levels, such as both middle and high school grades.  In our analyses, we will investigate whether the association between public schools and crime differs depending on the educational level of the school.  

We categorize each US census block in Philadelphia as either being near to a school or not near to a school.  Specifically, we define a US census block as being near to a school if any portion of that census block overlaps a 200m radius ($~1.5$ Philadelphia city blocks) around the GPS coordinates of that school location.  For the 18872 US census blocks in Philadelphia, this classification gives 5124 census blocks that are near to a school versus 13748 census blocks that are not near to a school.   In Figure S1 of our supplementary materials, we provide a map of Philadelphia that indicates the US census blocks that are near vs. not near to a public school.   

\subsection{Crime Data}

The Philadelphia Police department has made available data on every reported crime in Philadelphia from 2006 onwards.  Our analyses will be based on reported crimes from Jan 1, 2006 until January 19, 2020 which was when we acquired the data from the opendataphilly.org data portal.  For each reported crime, we have the date, time and GPS coordinates where the crime occurred as well as the type of each crime.  There are five broad categories of crimes:
\begin{enumerate}
\item {\bf Violent crimes}: aggravated assault, robbery, homicide, rape, other assaults, and other sex offenses.
\item {\bf Property crimes}: arson, burglary, motor vehicle theft, theft, and receiving stolen property.
\item {\bf Drug crimes}: narcotics
\item {\bf Other crimes}: disorderly conduct, driving under the influence, liquor law violations, offenses against family and children, prostitution, public drunkness, vagrancy, vandalism, and weapon violations
\end{enumerate}
There are a few additional crime types (e.g. embezzlement, forgery, fraud, and gambling) that are not considered in our analysis.  In Figure S2 of our supplementary materials, we provide maps of Philadelphia that compare the spatial distribution of total, violent, drug and property crimes.   We will focus our analyses on total crime incidence over the time period of our data, but we will also run some analyses separately for violent vs. property vs. drug crimes.  

\subsection{Economic and Demographic Data}

We have detailed data on neighborhood-level income, poverty, race, age and population density from the United States Census Bureau.  From the 2010 Census, we have the total population within every Census block as well as the percentage of white, black, asian, and hispanic residents, and the percentage of young residents (ages 15 to 29).  From the 2017 American Community Survey, we have the median household income and the number of families in poverty within every Census block group.  These datasets were downloaded using the Census API through the {\tt tidycensus} R library \citep{WalHer21}.

\subsection{Land Use Data}\label{landusedata}

The city of Philadelphia provides the zoning designations for the approximately half million lots in the city which we acquired from the opendataphilly.org data portal on Jan 19, 2020.   From this zoning data, we classified each lot in the city into one of nine land use categories: residential, commercial, industrial, civic, transport, culture, water, vacant, and other.  This data allows us to summarize the proportion each type of land use within any particular region of Philadelphia.   Although the geographical units for our analyses are US census blocks, we will use land use proportions that are calculated at the larger US census tract level in order to better represent the variation in land use in the larger area around each school.   

\subsection{Neighborhood Disorder Data}

The city of Philadelphia provides data on various indicators of neighborhood disorder, including public litter, 311 maintenance requests and code violations.  Litter data is made available by the City of Philadelphia and was downloaded on Jan 19, 2020 from the opendataphilly.org data portal.  This litter data consists of a {\it litter index} for every geocoded street segment in the city.  This litter index is a qualitative assessment made by city workers according to the amount of work necessary to remove the accumulated litter on the streets, and ranges in value from one to four with one the less work-intensive and four the most severe.  We calculate the litter index for each US census block by averaging the litter indexes for all street segments contained in that census block and across the two years, 2017 and 2018, that we have litter index data.  


Data on 311 maintenance requests and code violations in Philadelphia was also downloaded on Jan 19, 2020 from the opendataphilly.org data portal.  The code violations consists of all geocoded code violations from 2009 to 2020.   From this data, we calculate the total number of code violations for every US census block in Philadelphia.  We normalize these values by dividing the total number of code violations by the area of the census block and then log-tranforming them to create a {\it code violations index} for each census block. 

From the 311 maintenance requests data, we extracted six different variables: 1. Abandoned vehicle, 2. Streetlight out, 3. Graffiti, 4. Dumping, 5. Infestation, 6. Street defect.  For every US census block, we calculate the total number of occurrences of each of these six variables.   We normalize each of these six variables by dividing by the area of the census block and log-tranforming in order to create six different {\it 311 maintenance indices} for each census block. 

\subsection{Business Data}\label{businessdata}

Our research group manually assembled a database of Philadelphia business locations from three different web resources (Google Places, Yelp, and Foursquare). Each business is categorized into one or more of eight business types: Cafe, Convenience, Gym, Institution, Liquor, Lodging, Nightlife, Pharmacy, Restaurant, and Retail.  Additional details about this data is given in Humphrey et al. (2020).

From this database, we create measures of the presence of each business type around each US census block, which will be the geographical units of our analyses in Sections~\ref{eda}-\ref{matching}.   Specifically, we examined the average distance between the centroid of each census block and the nearest $k$ businesses of each business type.  We explored different values of $k$ between 1 and 10, and chose a particular value of $k$ for each business type based on which $k$ had the highest correlation with total crime which will be the primary outcome measure we examine in our analyses.  For most business types, we chose a value of $k$ between 3 and 5, except for $k=10$ for retail which is the most prevalant business category.  We will use these distances to $k$ businesses for each business type as our measures of business presence around each US census block.  

\section{Simple Comparisons of Crime near vs. not near to Public Schools}\label{eda}

Our overall goal is to investigate whether crime incidence is substantially different in local areas (or neighborhoods) of Philadelphia that are near to a public school compared to neighborhoods that are not near to a public school.   In most of our analyses, we will define these local areas as the US census blocks provided by the US Census Bureau.   As we discussed in Section~\ref{data}, we classified each of the US census blocks in the city of Philadelphia based on whether that census block overlaps a 200 meter radius around any public school location or not.   

\begin{figure}[ht!]
\includegraphics[width=15cm]{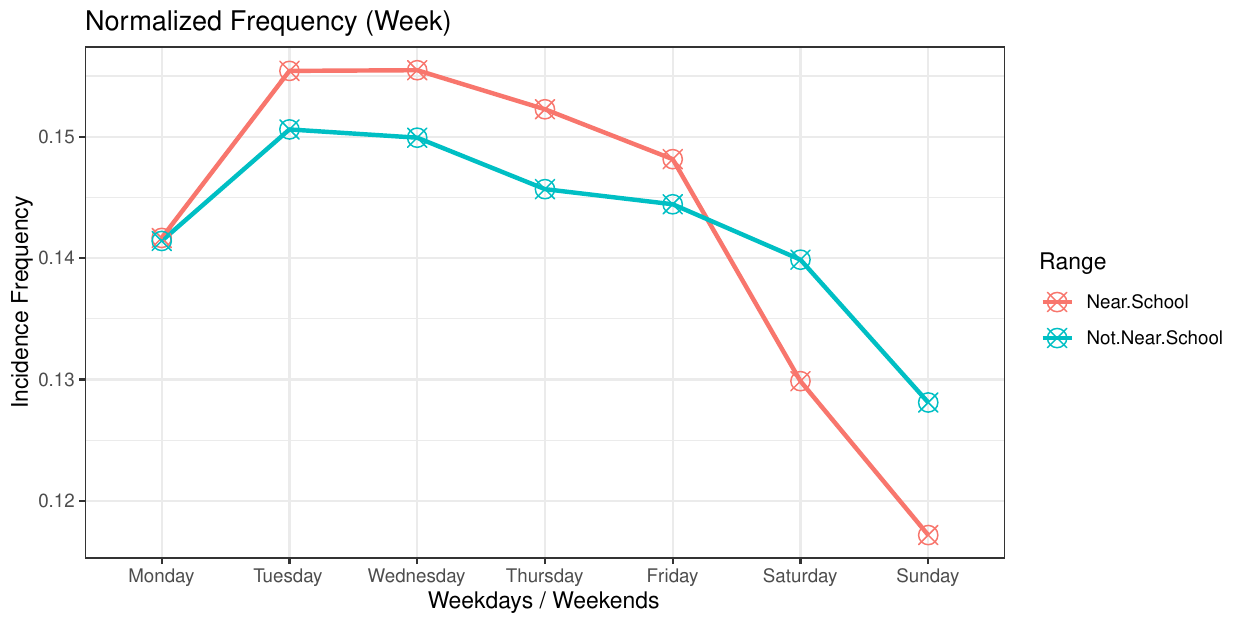}
\centering
\caption{Weekly trends in total crime incidence (per unit area) for US census blocks near to a public school (red) versus not near to a public school (blue)}
\label{crime-week}    
\end{figure}

We begin our analysis in Figure~\ref{crime-week} by comparing the weekly trends in total crime incidence (per unit area) between census blocks that are near versus not near to a public school.   We see a much greater difference between weekday and weekend crime incidence within the census blocks near to public schools, with elevated crime incidence during weekdays and lesser crime incidence during weekends compared to census blocks not near to public schools.  

\begin{figure}[ht!]
\includegraphics[width=14cm]{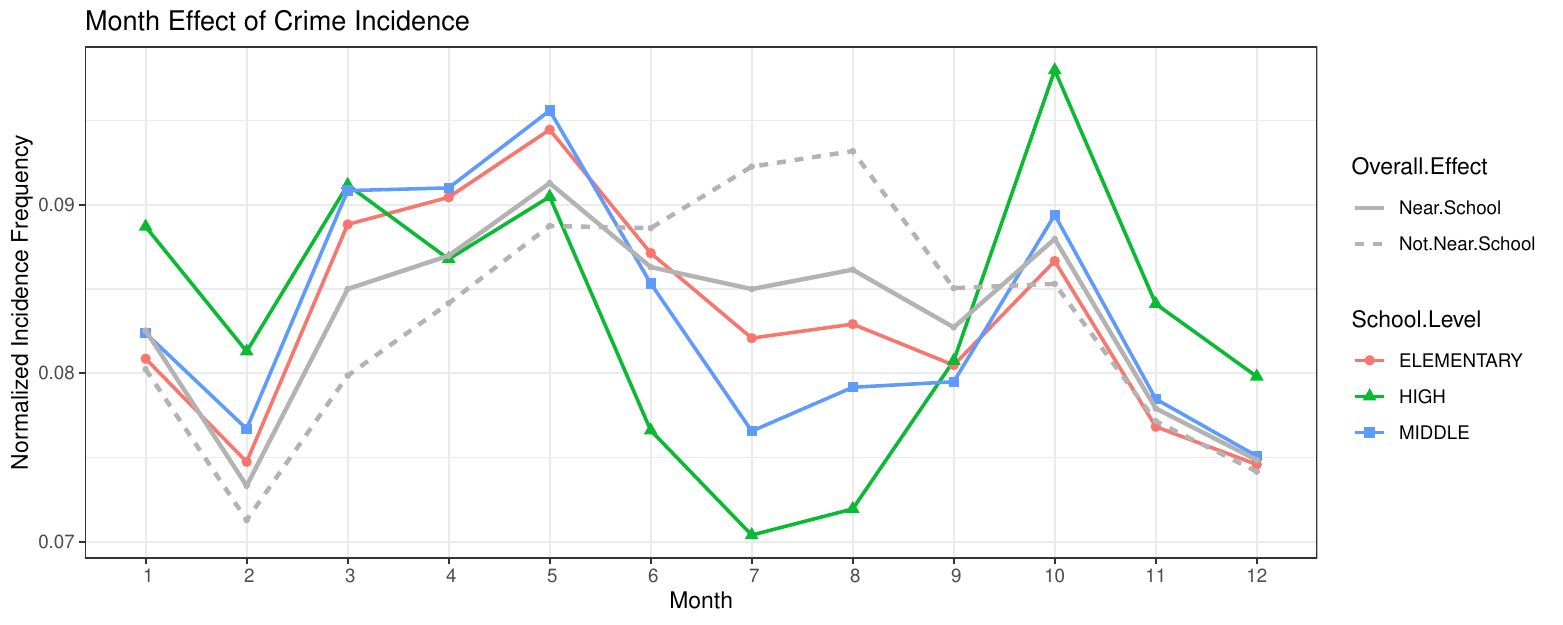}
\centering
\caption{Monthly trends in total crime incidence (per unit area) for US census blocks near to a public school (solid gray line) versus not near to a public school (dashed gray line).  Seperate trends (colors) given for census blocks near to public schools with different educational levels}
\label{crime-month}    
\end{figure}

We can also examine trends in total crime incidence throughout the year.  In Figure~\ref{crime-month}, we compare the monthly total crime incidence (per unit area) between census blocks that are near versus not near to a public school.  We see that the block groups near to a public school (solid gray line) have lower crime incidence during summer months (when school is not in session) compared to block groups not near to a public school (dashed gray line).  We also separated out these monthly trend depending on the educational level of the nearby school.  We see that the reduction in crime incidence during summer months is most dramatic for block groups near to public high schools with lesser reductions for middle and elementary schools. 

These weekly and monthly trends in crime incidence motivate us to divide the Jan 1, 2006 - Jan 19, 2020 time range covered by our available crime data into two distinct subsets based on when crime incidence is most likely to be affected by school proximity.   Specifically, we defined 6 am to 8 pm on weekdays from January to May and from September to December as our ``in-session" time period when students would be most likely to be in school or traveling/congregating in the area around their school.  Out of the full collection of $\approx$2.7 million crime incidences in our data, $\approx$1.1 crime incidences fall in these in-session time periods.  We label the time periods (and the $\approx$1.6 million crime incidences) outside of these target time windows as the ``out-of-session" subset of our crime data.  We will conduct (and compare) all of our analyses separately for these ``in-session" and ``out-of-session" subsets of time.  

We separately aggregate the number of total crimes, violent crimes, property crimes and drug crimes within each census block in Philadelphia and within the in-session vs. out-of-session time windows.   Total crime counts for each census block were divided by the area of that census block to generate a {\it total crime index} for each census block.   Our primary focus will be on the total crime index during the in-session time period but we will also perform some of our analyses on the out-of-session time period and separately for violent vs. property vs. drug crimes.  

\begin{figure}[ht!]
\includegraphics[width=15cm]{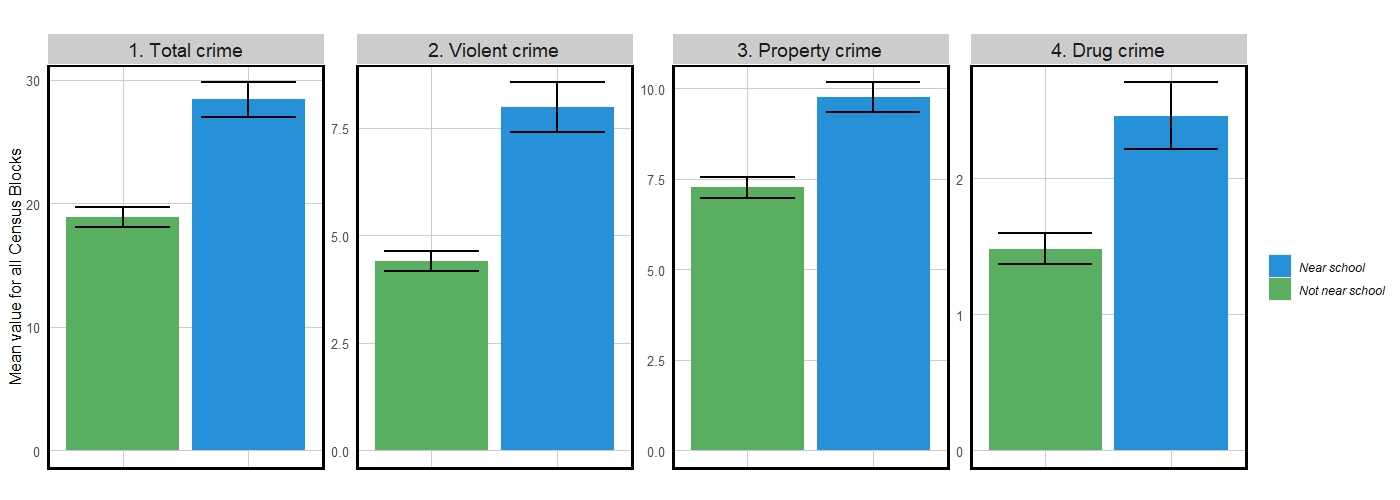}
\centering
\caption{Comparison of the average crime index within in-session time periods between Philadelphia census blocks near to a public school (blue) versus not near to a public school (red).  Separate bar plots are given for total crimes, violent crimes, property crimes and drug crimes. }
\label{crime-barplot}    
\end{figure}

In Figure~\ref{crime-barplot}, we compare the crime incidence during in-session time periods between Philadelphia census blocks that are near to public schools versus census blocks that are not near to public schools.  We see in Figure~\ref{crime-barplot} that, during in-session time periods, the average total crime index is substantially higher for census blocks near to public schools, as are the violent, drug and property crime indexes.   In Figure S3 of our supplementary materials, we provide the same comparison of crime incidence but for the out-of-session time periods and we find that crime incidence is higher near to public schools during out-of-session time periods as well.   From this simple comparison, we would conclude that crime incidence within Philadelphia is higher in the areas near to public schools.

\begin{figure}[ht!]
\includegraphics[width=15cm]{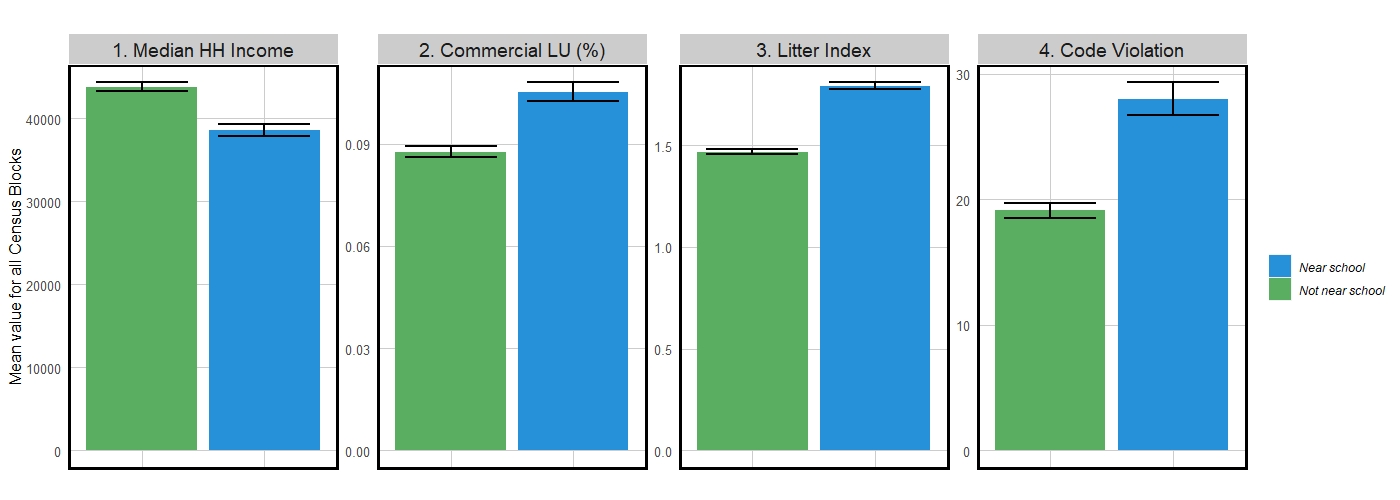}
\centering
\caption{Comparison of neighborhood characteristics between Philadelphia census blocks near to public schools versus not near to public schools.  From left to right, we compare 1. median household income, 2. percentage of commercial land use, 3. average litter index, and 4. average code violation index.}
\label{covariate-comparison}    
\end{figure}

However, we must be careful about interpreting these differences as the effects of school proximity on crime since the census blocks near to schools may differ from census blocks not near to schools on a myriad of other neighborhood characteristcs which are themselves associated with crime.  As examples, in Figure~\ref{covariate-comparison} we see substantial differences between census blocks near to versus not near to public schools in terms of their median household income, commercial land use, average litter index, and average number of code violations.  Census blocks near to schools have lower median household income, higher percentage of commercial land use and also have higher average litter and code violation indices.  Though not shown here, we also see differences on many of the other measures described in Section~\ref{data}.  

These substantial differences in surrounding neighborhood characteristics between the census blocks with vs. without public schools make it difficult to attribute any observed differences in crime incidence to the presence of a public school.  In Section~\ref{methods}, we outline two different statistical approaches that account for these differences: linear regression modeling (Section~\ref{regression}) and a propensity score matching analysis (Section~\ref{matching}). 

\section{Methodology to Account for Neighborhood Characteristics} \label{methods}

\subsection{Linear Regression Modeling} \label{regression}

In this linear regression approach, we consider total crime incidence within each census block in Philadelphia as our outcome variable.  We are specifically interested in whether being near to a public school is a significant predictor of total crime while controlling for other surrounding neighborhood characteristics that may also be predictive of total crime incidence.   We will estimate different effects for being near to public schools with different educational levels, i.e. elementary vs. middle vs. high public schools.  


We begin our linear regression modeling by noting that distribution of the total crime index created in Section~\ref{eda} across Philadelphia census blocks is quite skewed and contains some outlying values.  In Section 2 of our supplementary materials, we compare the distribution of the total crime index calculated over the in-session time window with the distribution of the logarithm of the total crime index and find that the log scale is a better fit to the standard linear regression assumption of normally distributed errors.

We consider the following linear model for the logarithm of the total crime index $y_{i}$ in Philadelphia census block $i$:  
\begin{eqnarray}
\log \left( y_{i} \right) = \beta_0 + \Bbeta \cdot \X_{i} + \phi \cdot S_{i} + \epsilon_{i} \label{regressioneqn}
\end{eqnarray}
where $\epsilon_{i} \sim {\rm Normal}(0,\sigma^2)$ and $\X_i$ are the demographic, economic, land use, disorder and business characteristics of census block $i$ as outlined in Section~\ref{data}.  Specifically, we included the following characteristics as potential predictors of crime: median household income, families in poverty, racial proportions, age brackets, land use proportions, distances to different business types, and indices for litter, 311 and code violations.   

The variable $S_{i}$ is an indicator variable where $S_{i} = 1$ if census block $i$ is near to a school, and $S_{i} = 0$ otherwise.  As outlined in Section~\ref{schooldata}, we define a census block as being near to a school if any portion of that census block overlaps a 200m radius around the GPS coordinates of that school location. 

The coefficient $\phi$ can be interpreted as the partial effect on the logarithm of total crime of a census block being near to a public school.  An  estimated $\phi$ that is significantly different from zero would suggest that proximity to a public school is significantly associated with total crime even after controlling for the effects ($\Bbeta$) of the surrounding neighborhood characteristics ($\X_{i}$).  

However, there is also the possibility that the effect of proximity to a public school on total crime is different depending on the educational level of that public school.  We address this possibility within our regression framework by estimating three additional versions of model (\ref{regressioneqn}) where the indicator variable $S_{i} = 1$ only if the census block $i$ is near to an elementary school vs. middle school vs. high school.  By comparing the estimated $\phi$'s from each of these four versions of (\ref{regressioneqn}), we can evaluate the association between total crime and proximity to elementary vs. middle vs. high vs. any public school.   

This regression framework also allows for other types of comparisons.  Beyond our examining different educational levels, we are also interested in whether the relationship between crime and proximity to public schools is different between the ``in-session" versus ``out-of-session" time periods defined in Section~\ref{eda}.   We also explored regression models where only violent crimes, property crimes or drug crimes were the outcome variable.
In Section~\ref{results}, we will focus on results for total crime as the outcome variable but we present results for the other combinations of crime outcomes and time periods in our supplementary materials.  In the next subsection, we describe an alternative analysis for the relationship between proximity to public schools and crime that uses propensity score matching to create balanced comparisons between neighborhoods that are near versus not near to public schools.  

\subsection{Propensity Score Matching Analysis} \label{matching}

In this approach, we create paired comparisons consisting of matched pairs of census blocks (one block with a public school and the other without a public school) that have highly similar surrounding neighborhood characteristics.  These matched pairs allow us to better isolate the relationship between crime and proximity to a school based on within-pair comparisons that are balanced on their surrounding demographic, economic, business, land use and disorder characteristics.

We create these matched pairs using a propensity score matching procedure \citep{RosRub83} implemented in the {\tt MatchIt} package \citep{HoImaKin11} in {\tt R}.   
The {\it propensity score} for a unit (census block) in our analysis is the estimated probability of that census block being near to a school based its surrounding neighborhood characteristics.   We estimate these propensity scores using a logistic regression model with the near school vs. not near school indicator as the outcome and the demographic, economic, land use, business and disorder measures for each census block as predictors.  

Two census blocks with highly similar demographic, economic, land use, business and disorder characteristics will have highly similar propensity scores.  For each census block that is near a school, we create a matched pair by finding a census block not near a school that has a highly similar propensity score.  Thus, within each matched pair we have an ``apples-to-apples" comparison of two census blocks that have highly similar surrounding neighborhood context but differ in terms of being near to a school vs. not near to a school.   A full list of the measures included in the propensity score estimation is given in Figure~\ref{matching-balance} where we evaluate the improved balance obtained by this matching procedure.  

We then use our created sets of matched pairs of census blocks to estimate the effect of school proximity on crime incidence.   Specifically, we calculate 
\begin{eqnarray}
D = {\rm mean} \left( \log y^{s}_{i} - \log y^{n}_{i} \right) \label{within-pair-difference}
\end{eqnarray} 
where, within each matched pair $i$,  $y^{s}_{i}$ is the total crime index for the census block near to a school and $y^{n}_{i}$ is the total crime index for the census block not near to a school.  

Similar to our regression analysis, we calculate the mean within-pair differences (\ref{within-pair-difference}) seperately for the ``in-session" versus ``out-of-session" time periods defined in Section~\ref{eda}.   In addition to our main comparison based on total crime incidence, we also calculate comparisons based only on violent crimes, property crimes or drug crimes.

\section{Results of Regression and Matching Analyses} \label{results}

\begin{figure}[ht!]
\includegraphics[width=14cm]{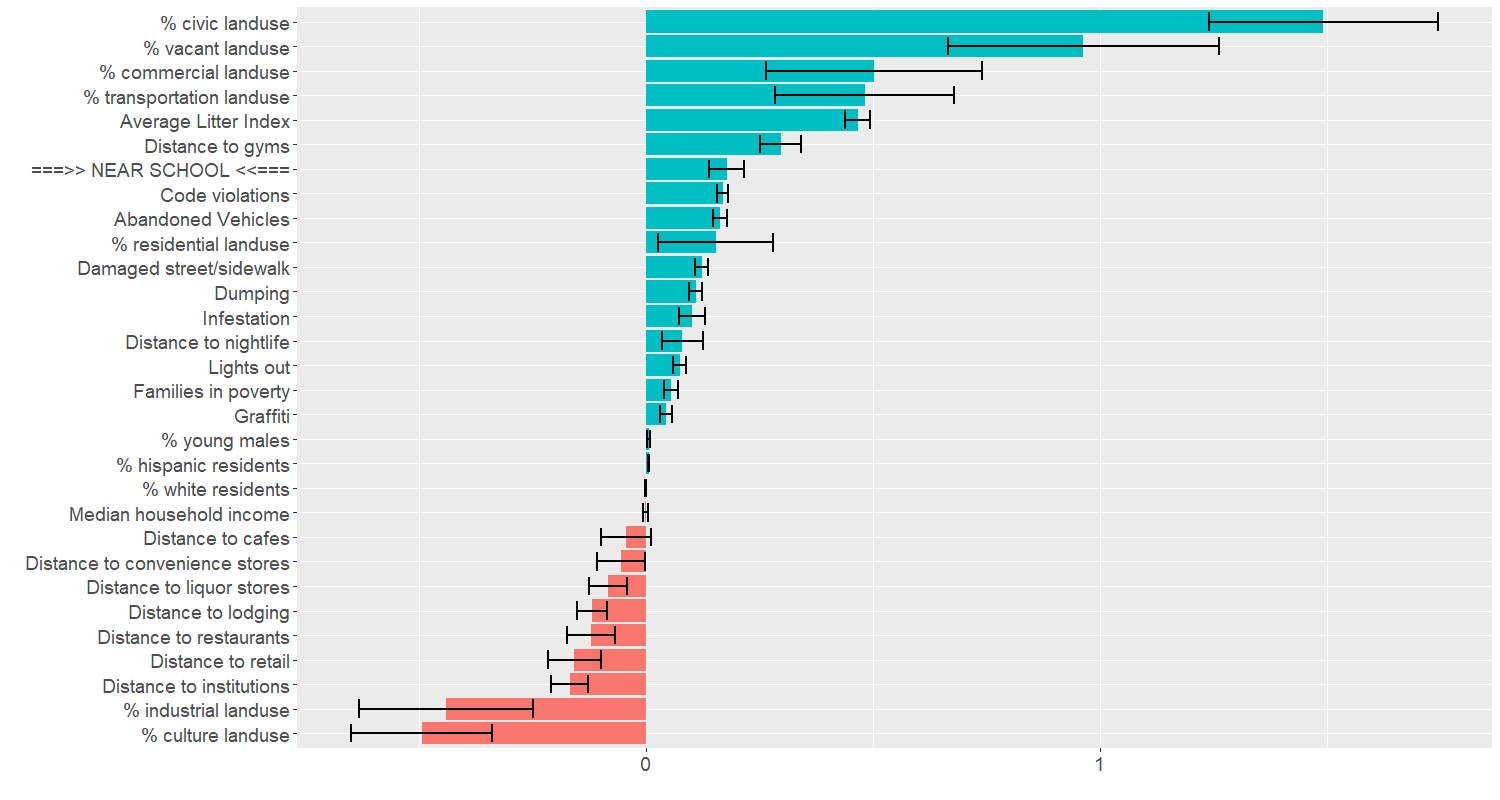}
\centering
\caption{Estimated coefficients ($\Bbeta$ and $\phi$) from linear regression model of total crime during ``in-session" time periods regressed on neighborhood characteristics $\X_{i}$  and an indicator variable $S_{i} = 1$ for proximity to any public school.  The effect of being near to a school ($\phi$) is emphasized with arrows and capital letters.}
\label{regression01-coef-barplot}    
\end{figure}

In Section~\ref{regression}, we described a regression model for estimating crime incidence around Philadelphia public schools, while accounting for the demographic, economic, business, land use and disorder characteristics of each neighborhood.    In Figure~\ref{regression01-coef-barplot}, we provide the estimated coefficients ($\Bbeta$ and $\phi$) for model (\ref{regressioneqn}) with total crime during ``in-session" time periods as the outcome variable and indicator variable $S_{i} = 1$ for proximity to {\it any} public school (regardless of educational level).  

We see in Figure~\ref{regression01-coef-barplot} that most of the largest (either positive or negative) partial effects on total crime at the census block level are the measures of land use zoning and business presence outlined in Sections~\ref{landusedata} and \ref{businessdata}.  Among the neighborhood disorder measures, the average litter index has the largest partial effect on total crime.  We are most interested in the partial effect $\phi$ of the indicator variable $S_{i}$ for proximity to any public school, which is labeled as ``NEAR SCHOOL" in Figure~\ref{regression01-coef-barplot}.  We see that being near to any public school has a significant positive partial effect on total crime.  

Since the outcome variable is log-transformed, we can interpret this partial effect as an expected percentage change in the outcome between $S_{i} = 0$ vs. $S_{i} = 1$.  A neighborhood near to a public school is associated with an expected increase of $\approx$ 18\% in total crime, holding all other variables in the model equal.  For a neighborhood with the median number of total crimes per block (89 total crimes per block over 14 years in Philadelphia), that 18\% corresponds to an extra 16 crimes (1.1 crimes per year).  

\begin{figure}[ht!]
\includegraphics[height=8cm,width=14cm]{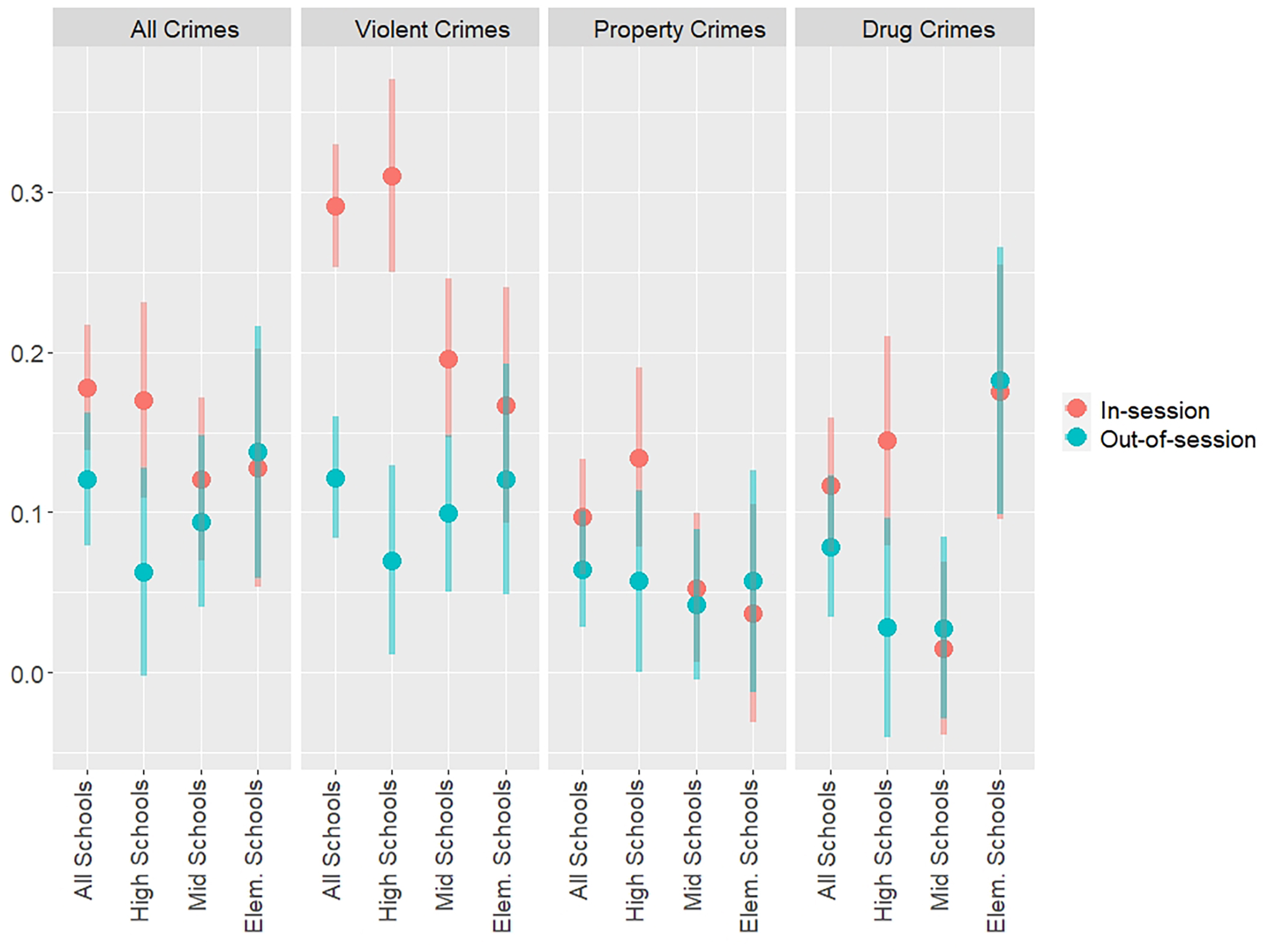}
\centering
\caption{Linear regression: comparing the estimated school effect $\phi$ between different crime outcomes, educational level of public schools, and time periods.  For each estimated $\phi$ we also provide vertical lines indicating $\pm \, 2 \cdot {\rm SE}(\phi)$. Vertical lines that do not overlap correspond to significant differences (at a $\alpha=0.05$ level).}
\label{schooleffect-comparison-reg}    
\end{figure}

In Figure~\ref{schooleffect-comparison-reg}, we compare the effect of being near a public school between different crime outcomes and educational level of the public school, as well as between the ``in-session" versus ``out-of-session" time periods.  We control for the same set of neighborhood characteristics $\X_{i}$ in each of these models. Overall, we see positive effects (i.e. increased crime incidence) of being near to a public school regardless of crime outcome, educational level and time period, but there is still substantial variation in the value of the estimated effect across these different conditions.  

In particular, we see substantially larger estimated school effects on violent crimes within the in-session time period compared to either property crimes or drug crimes.   Being near to any public school is associated with an expected increase of $\approx$ 29\% in violent crime in the in-session time period compared to an expected increase of $\approx$ 10\% in property crime.   For a neighborhood with the median number of violent and property crimes (20 violent and 31 property crimes per block over 14 years in Philadelphia), those percentages correspond to an extra 5.8 violent and 3.1 property crimes (0.4 violent and 0.2 property crimes per year).   

We also see substantial differences in the estimated school effect depending on the educational level of the school.  Across all four crime types within the ``in session" time period, we see substantially larger school effects for high schools compared to middle schools or elementary schools.  The difference is especially large for violent crimes in the in-session time period, where being near to a high school is associated with an expected increase of $\approx$ 31\% compared to $\approx$ 20\% for being near to a middle school and $\approx$ 19\% for being near to a middle school and $\approx$ 17\% for being near to an elementary school.   

Finally, we see generally larger estimated school effects on crime within the in-session time period compared to the out-of-session time period.  In particular, being near to any public school is associated with an increase of $\approx$ 18\% in total crime during the in-session time period compared to $\approx$ 12\% in the out-of-session time period, and is associated with an increase of $\approx$ 20\% in violent crime during the in-session time period compared to $\approx$ 12\% in the out-of-session time period.  

In estimating each of the school effects mentioned above, we are attempting to control for the surrounding neighborhood context by including the demographic, economic, land use, disorder and business measures outlined in Section~\ref{data} in our linear regression models.  However, there are possibly neighborhood aspects that may affect the relationship between school proximity and crime which are not captured by the variables we have included in our analyses, such as neighborhood watch groups, local community organizations, etc.  In our supplementary materials, we extend our regression model to allow for neighborhood-specific random school effects in order to explore additional neighborhood level variability in the relationship between crime and proximity to public schools.  

\begin{figure}[ht!]
\includegraphics[height=10cm,width=15cm]{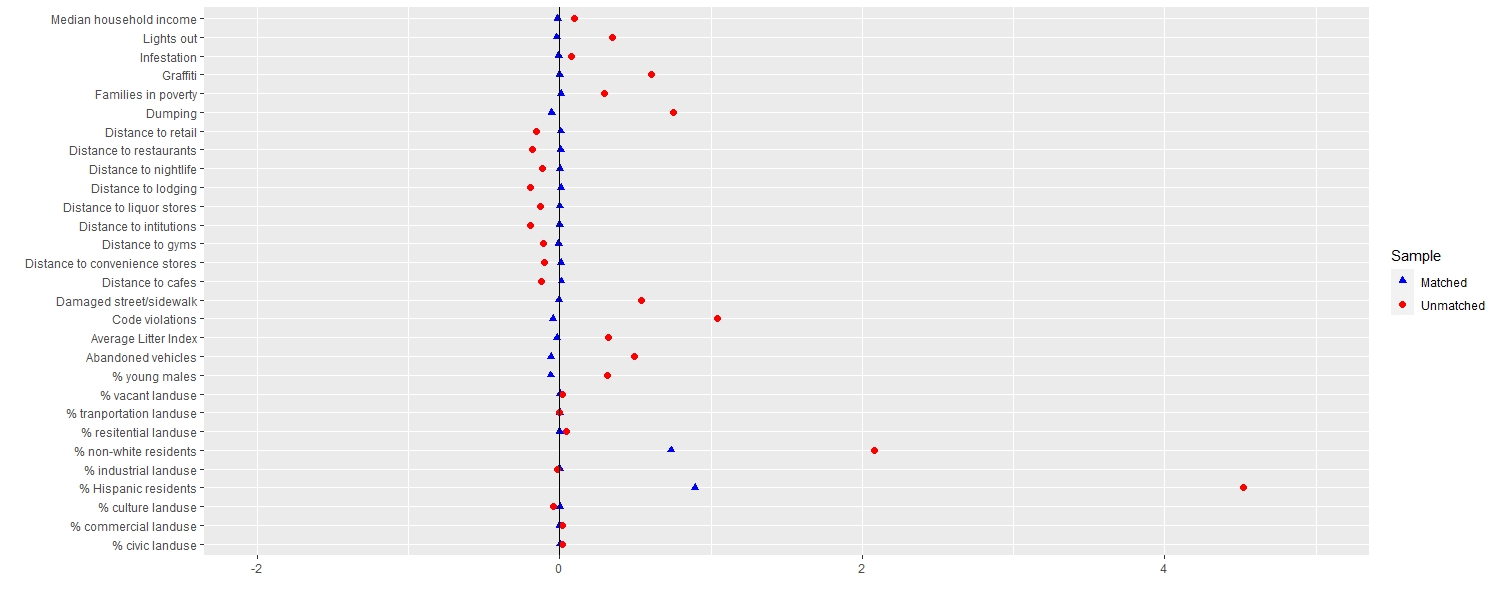}
\centering
\caption{Standardized differences for many surrounding neighborhood characteristics between census blocks near to a school versus not near to a school census.  For the ``matched" set, differences were calculated within the matched pairs created by our propensity score matching procedure. For the ``unmatched" set, differences were calculated between all unmatched census blocks near to a school versus not near to a school.}
\label{matching-balance}    
\end{figure}

As described in Section~\ref{matching}, matching analyses are an alternative approach for isolating the relationship between crime incidence and proximity to public schools from their surrounding neighborhood context.  In Figure~\ref{matching-balance}, we evaluate the improved balance on surrounding neighborhood characteristics that has resulted from our propensity score matching procedure.  Specifically, we calculate the standardized differences in each neighborhood characteristic within our matched pairs of census blocks near to a school and not near to a school and compare to these to standardized differences calculated between all unmatched census blocks near to a school versus not near to a school.  We see in Figure~\ref{matching-balance} that our matched pairs of census blocks have much less difference in their surrounding demographic, economic and land use characteristics which enables us to better isolate the relationship between crime incidence and proximity to a public school.  

\begin{figure}[ht!]
\includegraphics[height=10cm,width=15cm]{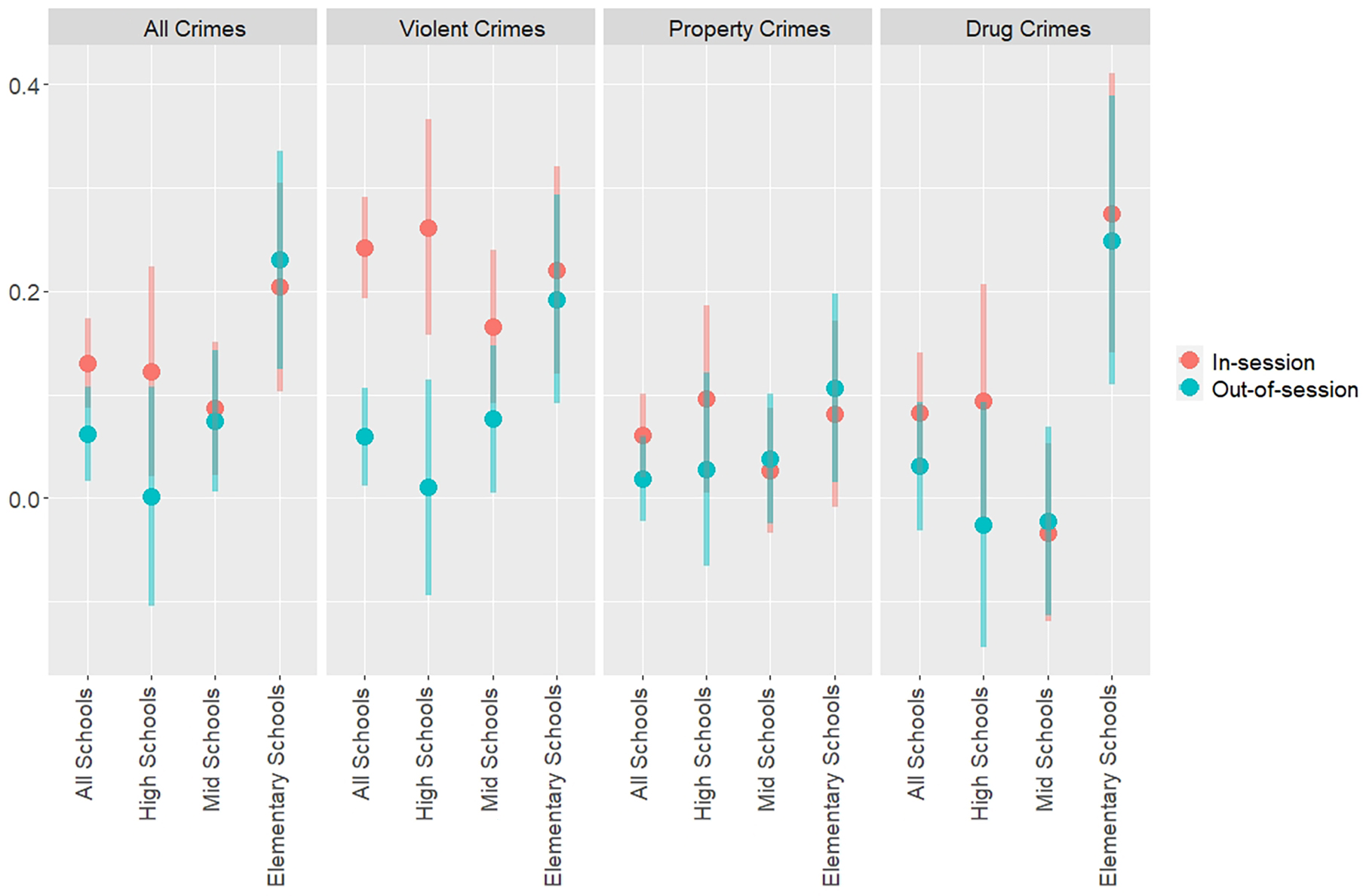}
\centering
\caption{Propensity score matching: comparing the estimated school effect $D$ within our matched pairs of blocks between different crime outcomes, educational level of public schools, and time periods.  For each estimated $D$ we also provide vertical lines indicating $\pm \, 2 \cdot {\rm SE}(D)$ where ${\rm SE}(D)$ is calculated using the paired version of the t.test package in the statistical software R. Vertical lines that do not overlap correspond to significant differences (at a $\alpha=0.05$ level).}
\label{matching-standdiff}    
\end{figure}

In Figure~\ref{matching-standdiff}, we compare mean within-pair differences (\ref{within-pair-difference}) between different crime outcomes and educational level of the public school, as well as between the ``in-session" versus ``out-of-session" time periods.   We also provide standard errors as vertical bars for each of these mean within-pair differences.  

We see very similar estimated effects of school proximity on crime in Figure~\ref{matching-standdiff} from these matched pair differences compared to the estimated effects in Figure~\ref{schooleffect-comparison-reg} from our regression modeling.   We do observe some subtle differences however, such as a larger positive effects of being near an elementary school on each crime type (as well as all crimes).   Similar to our regression analysis, we find a large effect on violent crimes in proximity to high schools during the in-session time period, where being near to a high school is associated with an increase of $\approx$ 27\% (compared to 31\% from our regression analysis).

Overall, this propensity score matching analysis confirms the results from our regression modeling in Section~\ref{regression}.   We find significantly increased crime incidence near to public schools regardless of crime outcome, educational level and time period, but with substantial variation in effects across these different conditions.  

\section{Discussion}\label{discussion}

In this paper, we explore the relationship between proximity to public schools and crime incidence from January 2006 to January 2020 in the city of Philadelphia.   In our analysis of this relationship, we must account for the surrounding economic, demographic and land use characteristics of locations with and locations without public schools.  We employ two statistical approaches, regression modeling and propensity score matching, in order to better isolate the association between crime and school proximity while controlling for surrounding neighborhood context.  


In both our regression modeling (Section~\ref{regression}) and propensity score matching (Section~\ref{matching}) approaches, we find significantly increased crime incidence near to public schools across almost all educational levels and time period.  However, the magnitude of elevated crime varies substantially between different crime outcomes, educational levels and time periods.  

We see the largest effects of school proximity on crime for the in-session time period, violent crime type, and high school educational level.  Our regression analysis finds that proximity to a high school in the in-session time period is associated with an expected increase of $\approx$ 31\% in violent crimes compared to $\approx$ 14\% in property crime and $\approx$ 17\% in total crime.   Similarly large effects were also found in our propensity score analysis, where proximity to a high school in the in-session time period is associated with an expected increase of $\approx$ 27\% in violent crimes compared to $\approx$ 15\% in property crime and $\approx$ 16\% in total crime. 

As we hypothesized, crime incidence around public schools is substantially higher during in-session time periods compared to out-of-session time periods for most crime types and education levels.  These results support the routine activities theory of \cite{CohFel79} and pattern theory of \citep{BraBra93}.   However, crime incidence around public schools is also significantly elevated (though with a smaller magnitude) during the out-of-session time periods for most crime types and education levels, which was not expected.  We also find that high schools have the highest elevated crime incidence during in-session periods among the education levels, which we hypothesized based on the peak ages of crime propensity \citep{Far86}.  

Finally, we hypothesized based on pattern theory \citep{BraBra93} that crime types could be differentially impacted by proximity to a school, and we observe substantial differences between our estimated violent crime effects versus property crime effects.  Specifically, we find that violent crime incidence is more elevated during in-session periods compared to out-of-session time periods (especially around high schools), whereas property crime incidence is not substantially different between in-session and out-of-session time periods. 

We can also compare our results to \cite{SteUkeMac19} where crime incidence was examined following closures of academically under-performing public schools in Philadelphia.   \cite{SteUkeMac19} estimated a 15\% decline in total crime and a 30\% reduction in violent crime in the surrounding area following school closures, but did not find any effect on property crimes.  Similarily, we find total crimes to be elevated by around 14-16\% and violent crimes to be elevated by around 24-29\% around public schools during in session time periods.  However, we also find that property crimes are significantly elevated by around 7-10\% around public schools during in session time periods.

This comparison is especially interesting given that we evaluated crime incidence in the areas around all public schools in Philadelphia whereas \cite{SteUkeMac19} focussed only on a small subset of academically under-performing schools that were recently closed.  We generalize their results with our finding that elevated incidence of total and violent crimes is characteristic of the areas around all public schools rather than just the worst performing schools in the city.   In contrast with \cite{SteUkeMac19}, we also find significantly elevated incidence of property crimes around public schools in Philadelphia but we see very little separation in property crime incidence between in-session and out-of-session time periods, especially compared to the large differences (especially around high schools) in violent crime incidence that we find between in-session and out-of-session time periods.  


Beyond our focus on the presence of public schools, our regression approach also provides insight into other neighborhood features that are associated with crime incidence.  Commercial land uses are highly correlated with crime incidence which was also found by \cite{BerBlo09}.  The litter index is also highly correlated with crime which supports the ``Broken Windows" theory of \cite{WilKel03}.  

We should also note that our results are based on the Philadelphia Police Department database of all {\it reported} crimes in Philadelphia which is not a complete census of all crimes that were committed in Philadelphia over the time span of our study.  \cite{BuiMorLan22} investigate the potential biases in police crime data at different levels of spatial aggregation.   To the extent that reported crimes are a biased representation of all committed crimes, some part of the elevated reported crime incidence found nearby to public schools could be driven by increased police attention and activity in those areas.   

Beyond potential biases in the crime data, the complex relationship between schools and public safety is presumably influenced by many other factors (e.g. home ownership and residential turnover) that we have not taken into account in this current study.  Particularly interesting would be measures of human activity and guardianship at different times of the day in order to provide a more detailed picture of the activity backcloth \citep{BraBra93} around each school. 

\section*{Acknowledgements} 

We are grateful to the Wharton Social Impact Initiative for their generous support of our work.  





\section*{Funding}\label{funding}

This work was supported by the Wharton Social Impact Initiative through the grant `Measuring Human Vibrancy and its Role in the Urban Development of Philadelphia" (2018).

\section*{Data Availability}\label{dataavailibility}

The github repository for our code and data can be accessed at: 
\begin{center}
\url{https://github.com/leonardoharth/School_Crime_Research}
\end{center}







\bibliographystyle{natbib}
\bibliography{references}  

\newpage

\begin{center}
{\bf Supplementary Materials for ``Spatial Analysis of the Association between School Proximity and Crime in Philadelphia"}
\end{center}

\renewcommand{\thesection}{S \arabic{section}}
\setcounter{section}{0}

\section{Visualizations of School and Crime Data}

\begin{figure}[ht!]
\renewcommand\thefigure{S1}
\includegraphics[width=12cm]{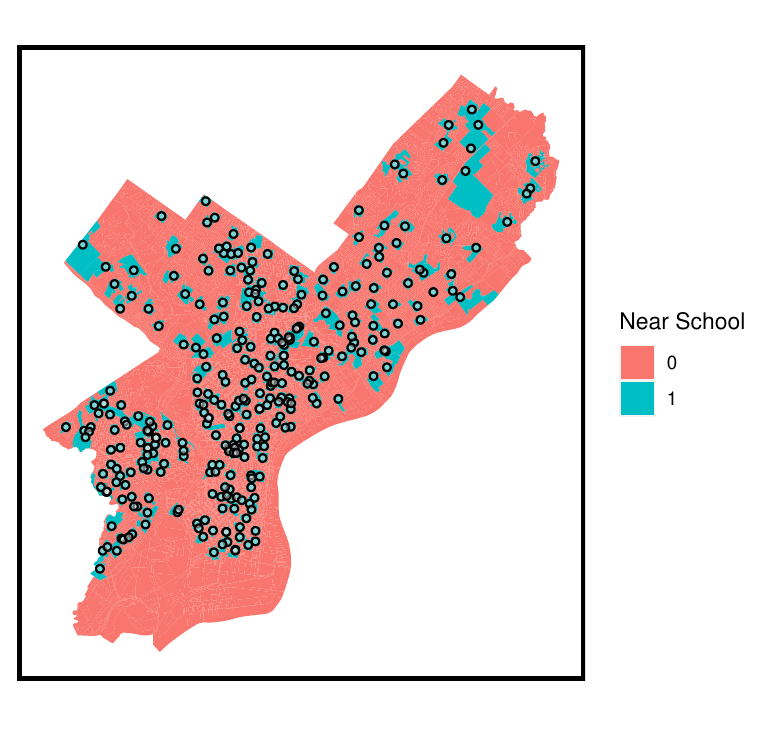}
\centering
\caption{Map of Philadelphia that indicates all schools (black circles) as well as all US census blocks near to those schools (light blue = 1) versus US census blocks not near to schools (red = 0).}
\label{map-blocks-near-schools}    
\end{figure}

Figure~\ref{map-blocks-near-schools} is a map of Philadelphia that indicates all US census blocks that are near to a school versus not near to schools.  As described in our main paper, we defined any US census block as being near to a school if any portion of that census block overlaps a 200m radius around a school location.

The four maps in Figure~\ref{map-crimes-comparison} below show the spatial distribution of total crimes, violent crimes, drug crimes and property crimes (all on the log scale). It is worth noting that drug crimes are more prone to spatial clustering than other types of crime in Philadelphia.

\begin{figure}[ht!]
\renewcommand\thefigure{S2}
\includegraphics[width=\textwidth]{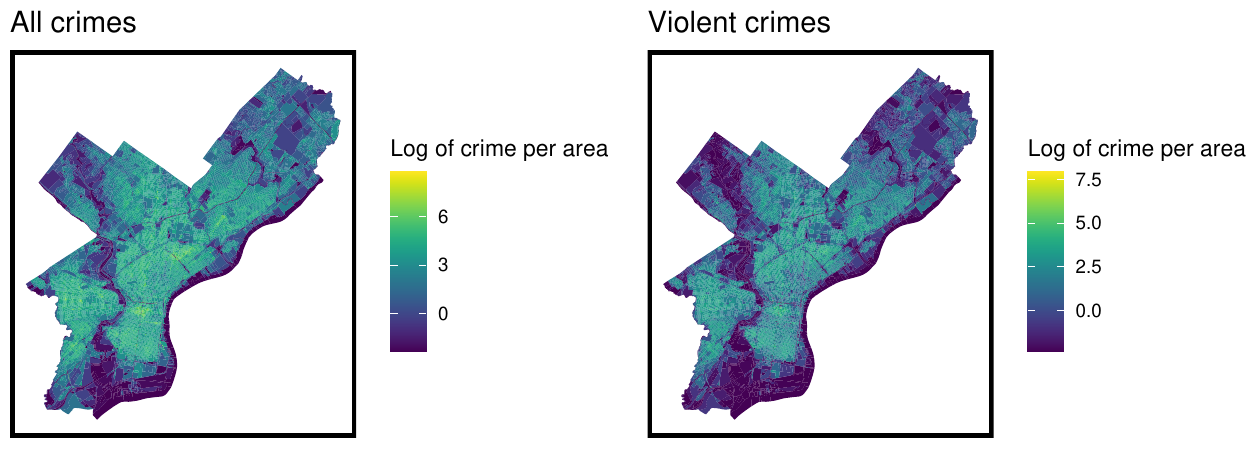} \\
\includegraphics[width=\textwidth]{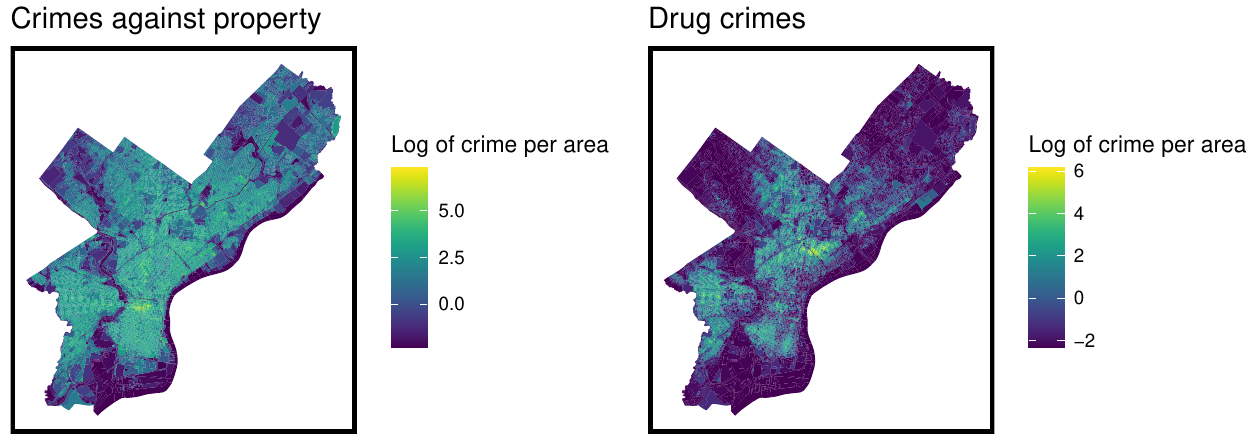} \\
\centering
\caption{Maps of Philadelphia that give the spatial distribution of total crimes, violent crimes, drug crimes and property crimes, all on the log scale}
\label{map-crimes-comparison}
\end{figure}

\begin{figure}[ht!]
\renewcommand\thefigure{S3}
\includegraphics[width=12cm]{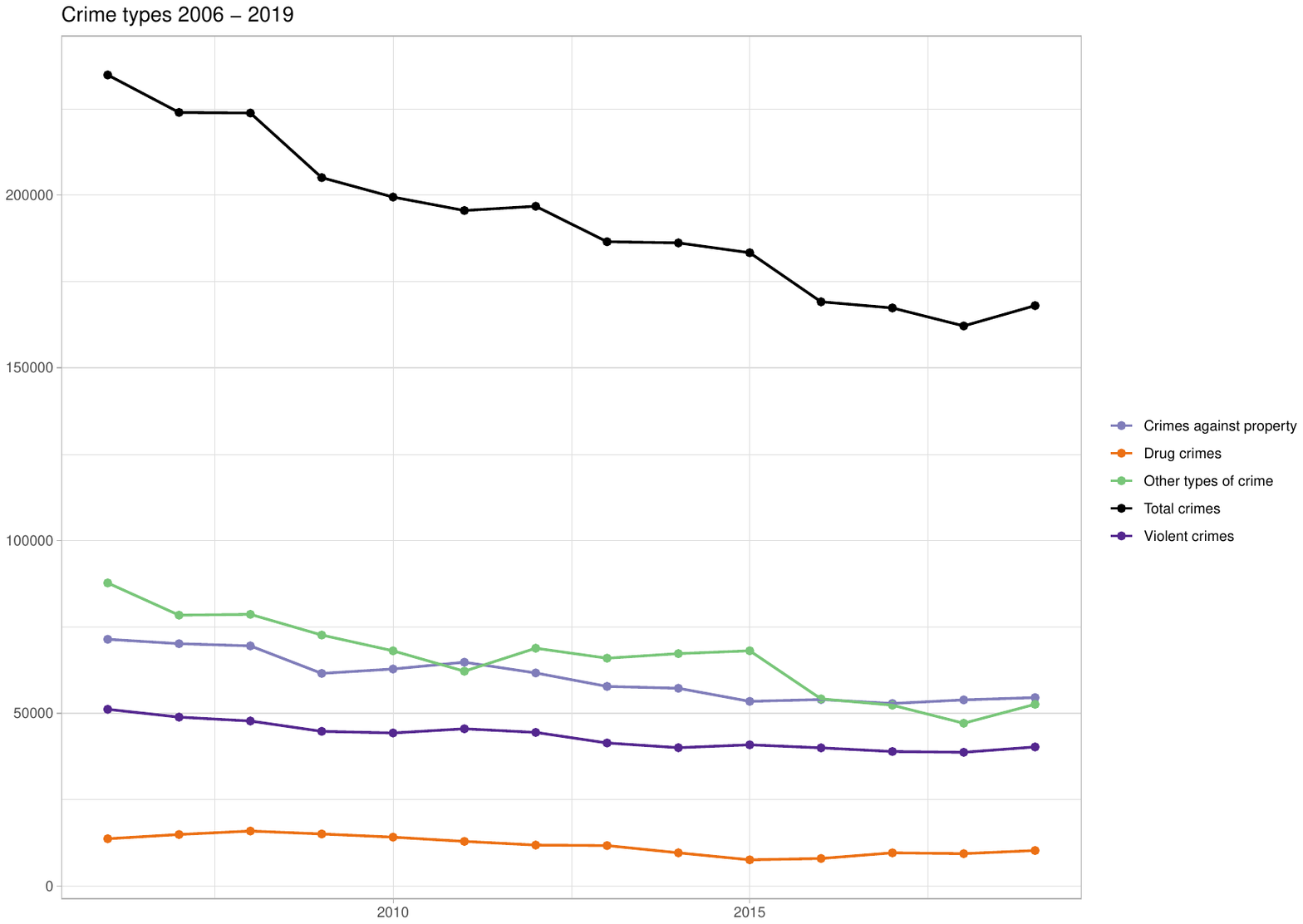}
\centering
\caption{Number of crimes per year of different types (all vs. violent vs. property vs. drug vs. other) for each year in our data}.
\label{crimes-over-time}    
\end{figure}

Figure~\ref{crimes-over-time} gives the trends over time for total crimes, violent crimes, drug crimes and property crimes.  We see that there is a small decrease in each crime type in Philadelphia from 2006 to 2019.  

\section{Comparison of Crimes in Philadelphia}

\begin{figure}[ht!]
\renewcommand\thefigure{S4}
\includegraphics[width=15cm]{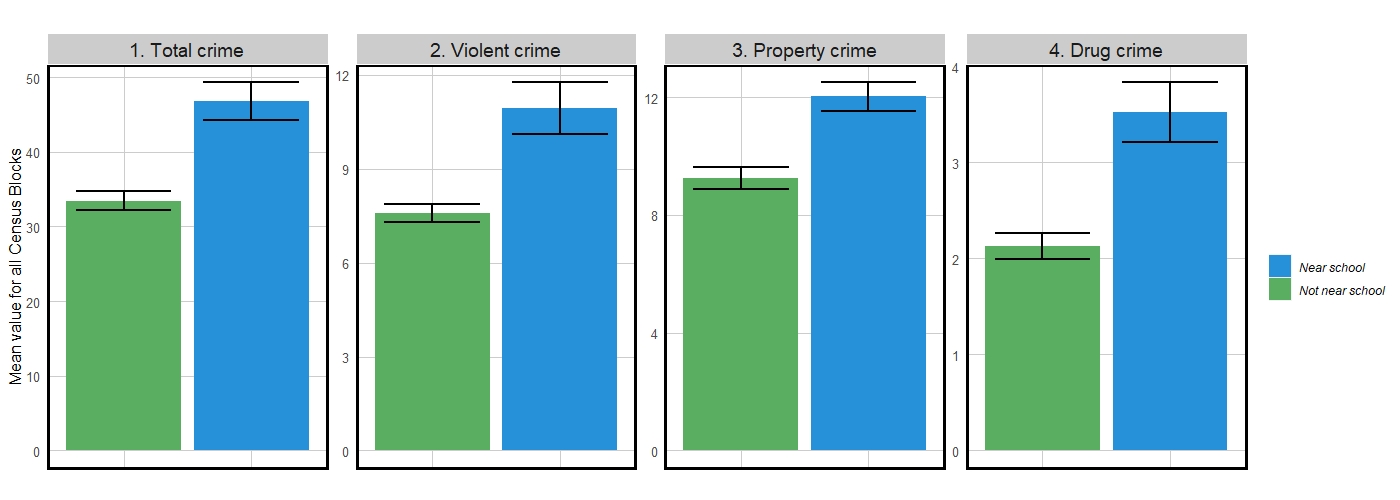}
\centering
\caption{Comparison of the average crime index within out-of-session time periods between Philadelphia census blocks near to a public school (blue)versus not near to a public school (red).  Separate bar plots are given for total crimes, violent crimes, drug crimes and property crimes.}
\label{crime-barplot-mirror}    
\end{figure}

In Figure~\ref{crime-barplot-mirror}, we compare the crime incidence during out-of-session time periods between Philadelphia census blocks that are near to public schools versus census blocks that are not near to public schools.  We see that, during out-of-session time periods, the average total crime index is substantially higher for census blocks near to public schools, as are the violent, drug and property crime indexes.   These results echo the same findings in Figure 3 of our main paper for crime incidence in the in-session time periods. 

\begin{figure}[ht!]
\renewcommand\thefigure{S5}
\includegraphics[width=\textwidth]{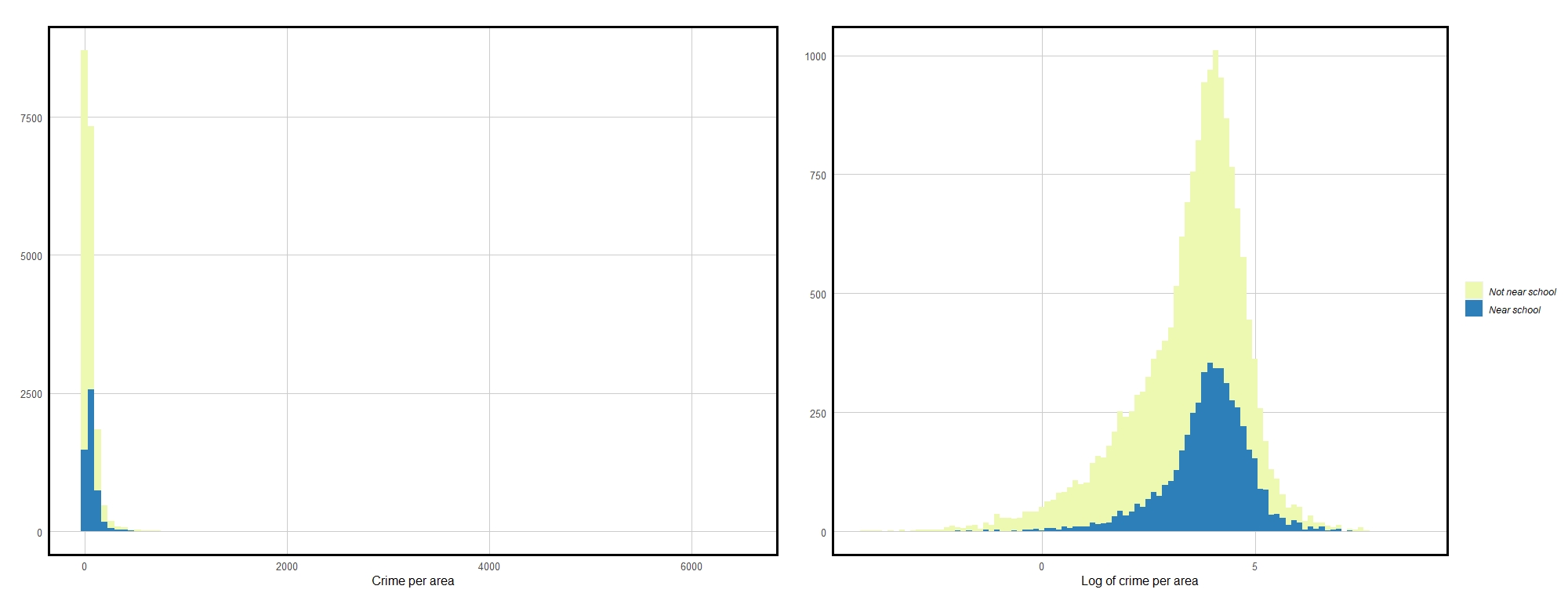}
\centering
\caption{Distribution of the total crime index in the target time window over all census blocks in Philadelphia on both the original scale (left) and log scale (right).  The blue histogram is census blocks near to a school whereas the green histogram is census blocks not near to a school.}
\label{crime-distribution}
\end{figure}

Figure~\ref{crime-distribution} gives the distribution of the total crime index in the target time window over all census blocks in Philadelphia on both the original scale (left) and log scale (right).  Separate histograms are given for census blocks near to a public school (blue) versus not near to a public school (red).  Note that a value of 0.1 was added to the crime index of each observation to avoid having to take the logarithm of values equal to zero.

\section{Random Effects Linear Regression Models} \label{randomeffects}

The regression model in Section 4 of our manuscript estimates effects of school proximity and other neighborhood characteristics that are {\it fixed}, i.e. shared by all neighborhoods in the city of Philadelphia.  However, as discussed above, there may be idiosyncratic neighborhood-specific effects of school proximity on crime that vary between different parts of the city, e.g. neighborhood watch groups.  

One way to address this possibility within our regression framework is by allowing different {\it random} effects of school proximity on crime for each neighborhood, while still estimating fixed effects for the other neighborhood characteristics in our model, i.e. 
\begin{eqnarray}
\log \left( y_{i} \right) = \beta_0 + \Bbeta \cdot \X_{i} + \phi_i \cdot S_{i} + \epsilon_{i} \label{regressioneqn2}
\end{eqnarray}
where $\epsilon_{i} \sim {\rm Normal}(0,\sigma^2)$.  

Comparing (\ref{regressioneqn2}) to the regression model in Section 4 of our manuscript, the only difference is that being near to a school ($S_{i}$ = 1) now has potentially different effects $\phi_i$ for different neighborhoods in (\ref{regressioneqn2}).  These neighborhood-specific school effects are random effects that are assumed to share a common distribution, $\phi_i \sim {\rm Normal}(0,\tau^2)$, whereas the effects $\Bbeta$ of the other neighborhood characteristics on crime still considered fixed and shared across all neighborhoods.

However, this new random effects approach also differs from our fixed effect approach in Section 4 of our manuscript in terms of spatial resolution of our analysis.   Allowing neighborhood-specific random effects to our model would lead to an extra $\approx 10000$ extra parameters in (\ref{regressioneqn2}) if we define our neighborhoods by US census blocks as we did in Section 4 of our manuscript.  A model with that many extra parameters would probably be quite unstable even with the regularization imposed by the shared prior distribution on those added random effects.  

So instead, we implemented neighborhood-specific random effects for each US census {\it block group} in Philadelphia, which only adds $\approx 1000$ extra parameters to the model.   We also changed our two comparison time periods to be simply weekdays vs. weekends, rather than the ``in-session" vs. ``out-of-session" time periods from Section 4 of our manuscript.  

\begin{figure}[ht!]
\renewcommand\thefigure{S6}
\includegraphics[width=14cm]{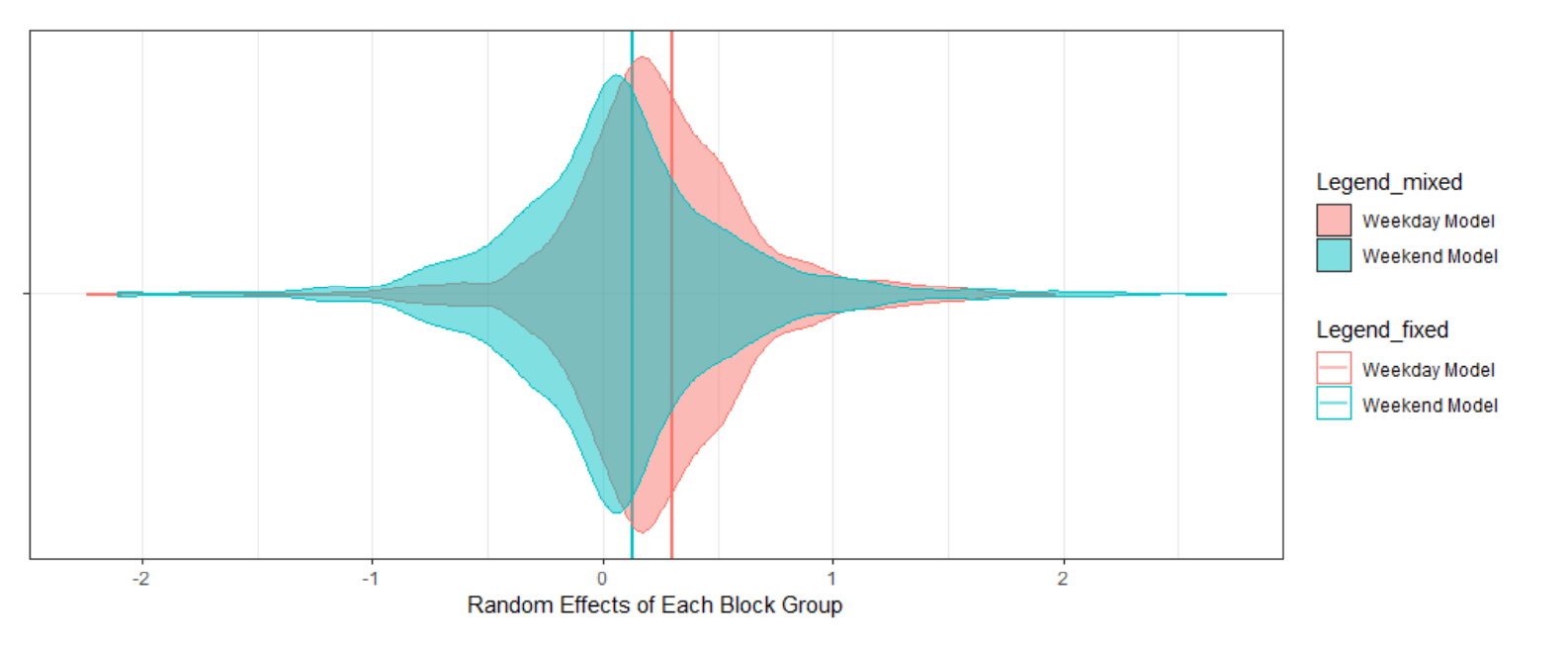}
\centering
\caption{Violin plots of the neighborhood specific effects $\hat{\phi}_i$ from model (\ref{regressioneqn2}), estimated using only weekday time periods (red) or only weekend time periods (blue).  Vertical lines also indicate the global effect of school proximity on total crime $\hat{\phi}$ from the fixed effect model, estimated using only weekday time periods (red) or only weekend time periods (blue). }
\label{fig-random-effects-violin}    
\end{figure}

Having neighborhood specific random effects allows us to visualize the variation between different neighborhoods in terms of their specific effects $\phi_i$ of being near to a public school on total crime.  In Figure~\ref{fig-random-effects-violin}, we compare the distribution of the estimated neighborhood specific effects of being near to a public school on total crime between the model (\ref{regressioneqn2}) estimated using just weekdays versus just weekends. 

We see quite a bit of variation between block groups in terms of their estimated random effects $\hat{\phi}_i$, which suggests that the relationship between school proximity and total crime does vary between different neighborhoods in Philadelphia.  We also see a substantial shift towards larger positive values in the distribution of random effects for weekdays (red) compared to weekends (blue).  These results provide additional support for the significantly positive effects of school proximity on total crime found in Section 4 of our manuscript.  

\section{Regression Tables}

\begin{figure}[ht!]
\renewcommand\thefigure{S7}
\includegraphics[width=\textwidth]{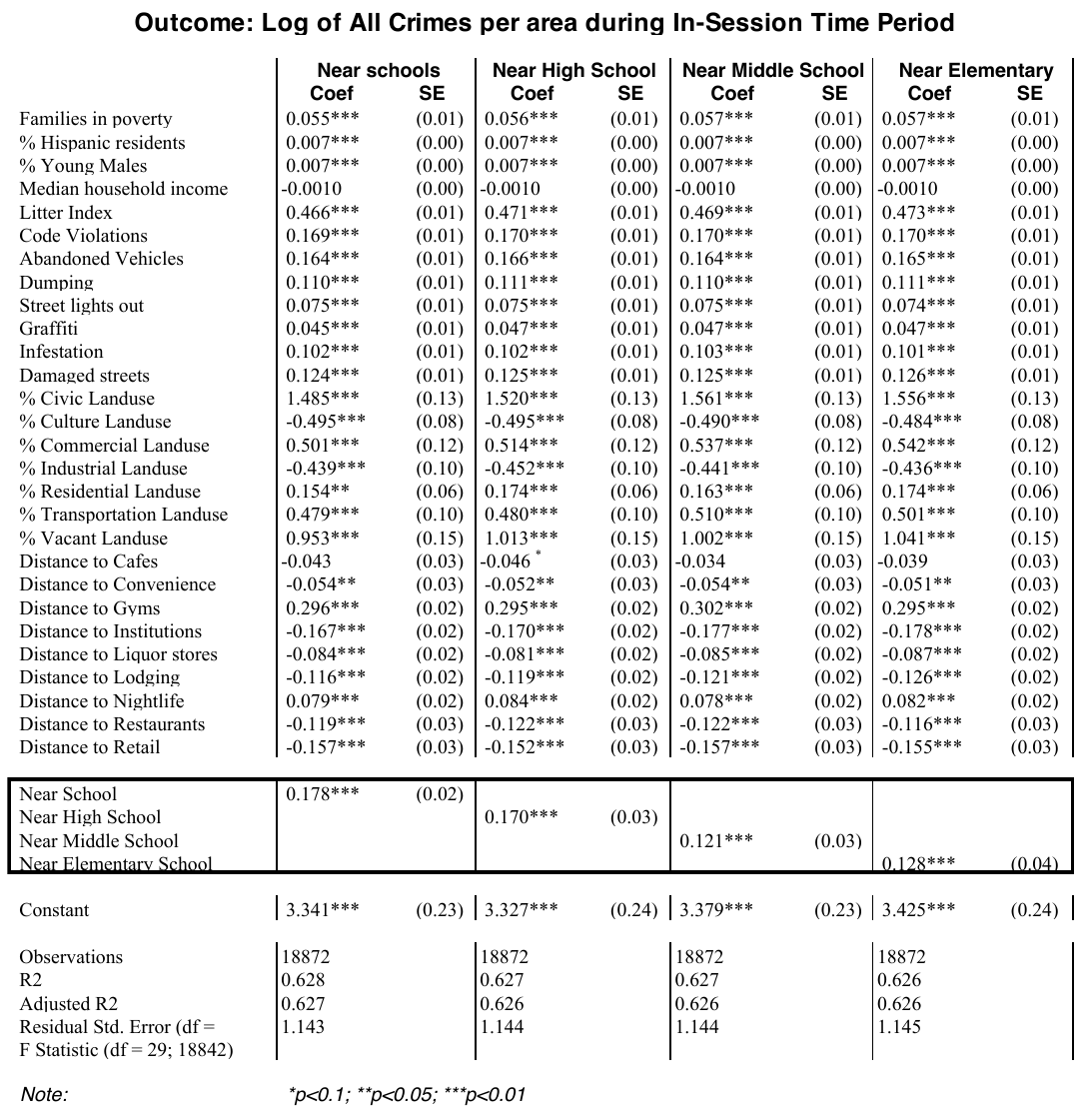}
\centering
\caption{Regression results with log of all crimes per area during in-session time periods as outcome variable}
\label{reg-results-01-04-all-insession}
\end{figure}

Figure \ref{reg-results-01-04-all-insession} gives results for the regression models with log of all crimes per area during in-session time periods as outcome variable.    

\begin{figure}[ht!]
\renewcommand\thefigure{S8}
\includegraphics[width=\textwidth]{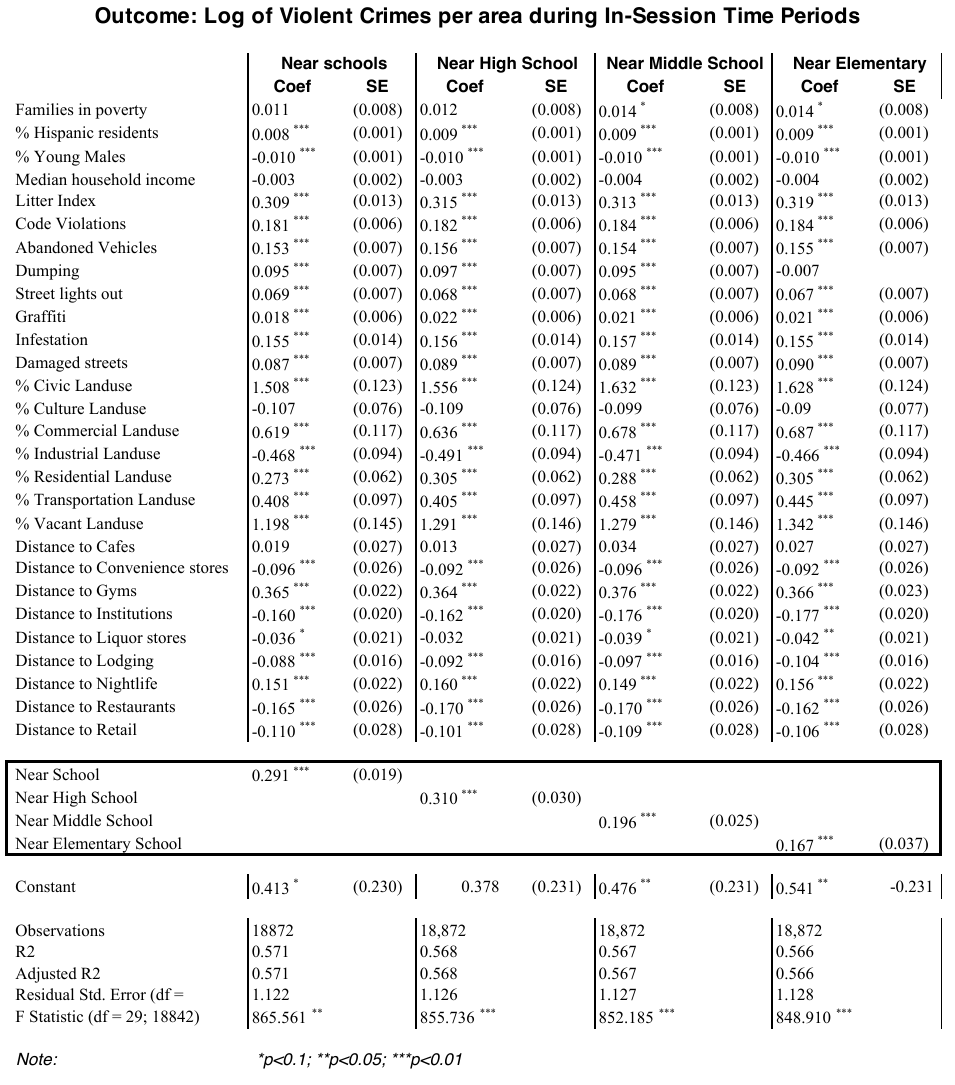}
\centering
\caption{Regression results with log of violent crimes per area during in-session time periods as outcome variable}
\label{reg-results-05-08-violent-insession}
\end{figure}

Figure \ref{reg-results-05-08-violent-insession} gives results for the regression models with log of violent crimes per area during in-session time periods as outcome variable.    

\begin{figure}[ht!]
\renewcommand\thefigure{S9}
\includegraphics[width=\textwidth]{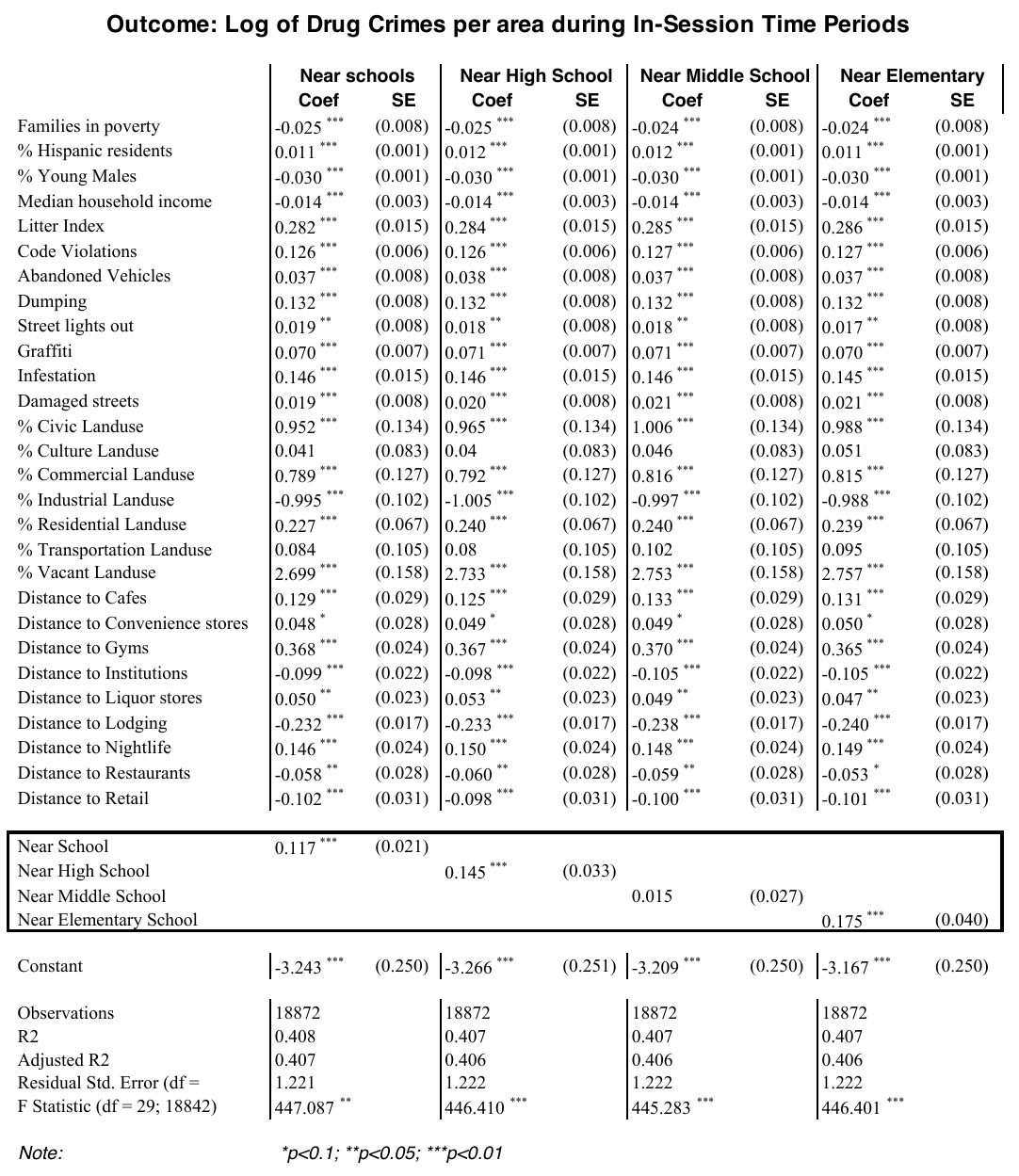}
\centering
\caption{Regression results with log of drug crimes per area during in-session time periods as outcome variable}
\label{reg-results-09-12-drug-insession}
\end{figure}

Figure \ref{reg-results-09-12-drug-insession} gives results for the regression models with log of drug crimes per area during in-session time periods as outcome variable.  

\begin{figure}[ht!]
\renewcommand\thefigure{S10}
\includegraphics[width=\textwidth]{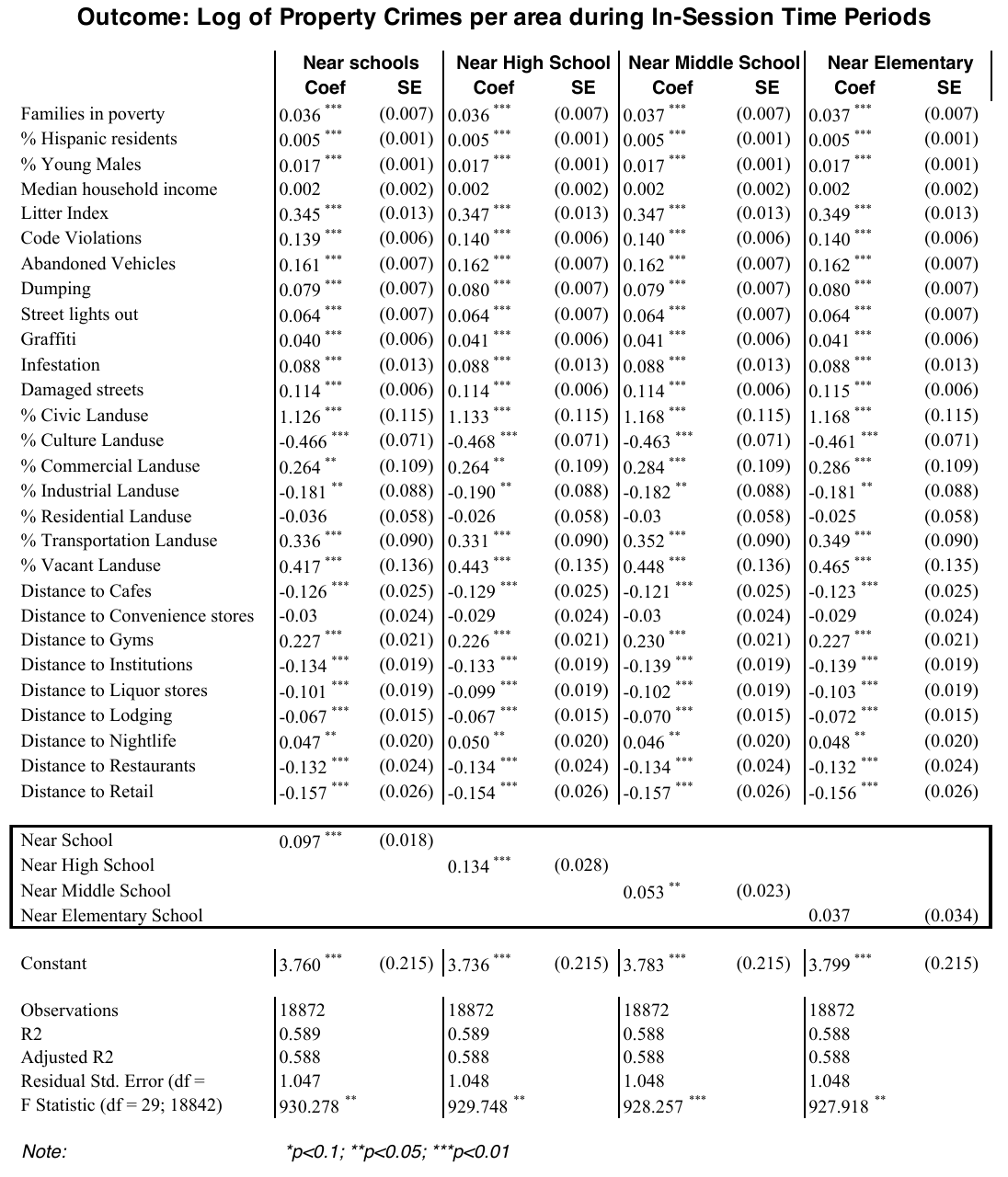}
\centering
\caption{Regression results with log of property crimes per area during in-session time periods as outcome variable}
\label{reg-results-13-16-property-insession}
\end{figure}

Figure \ref{reg-results-13-16-property-insession} gives results for the regression models with log of property crimes per area during in-session time periods as outcome variable.  

\begin{figure}[ht!]
\renewcommand\thefigure{S11}
\includegraphics[width=\textwidth]{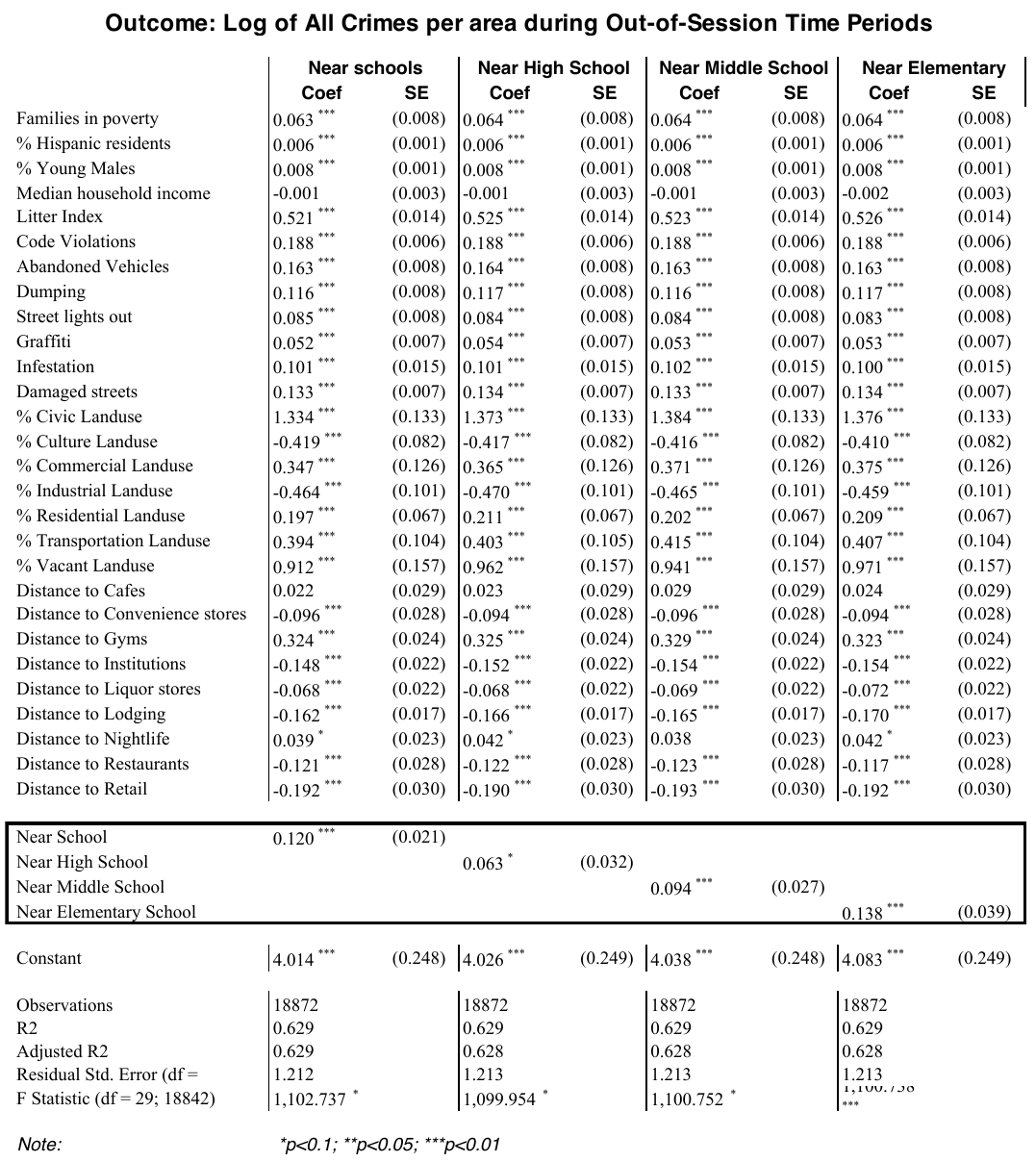}
\centering
\caption{Regression results with log of all crimes per area during out-of-session time periods as outcome variable}
\label{reg-results-17-20-all-outofsession}
\end{figure}

Figure \ref{reg-results-17-20-all-outofsession} gives results for the regression models with log of all crimes per area during out-of-session time periods as outcome variable.    

\begin{figure}[ht!]
\renewcommand\thefigure{S12}
\includegraphics[width=\textwidth]{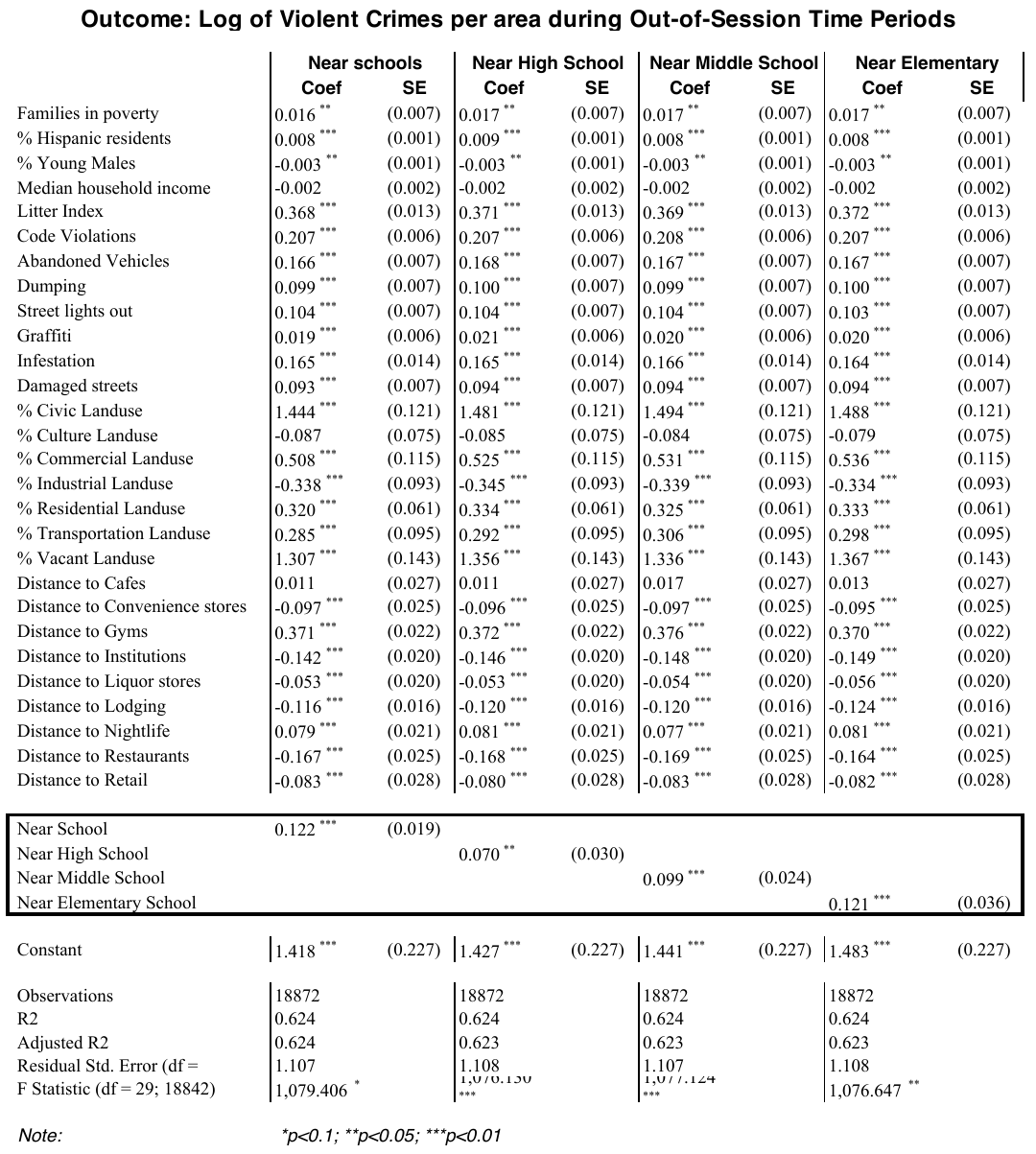}
\centering
\caption{Regression results with log of violent crimes per area during out-of-session time periods as outcome variable}
\label{reg-results-21-24-violent-outofsession}
\end{figure}

Figure \ref{reg-results-21-24-violent-outofsession} gives results for the regression models with log of violent crimes per area during out-of-session time periods as outcome variable.    

\begin{figure}[ht!]
\renewcommand\thefigure{S13}
\includegraphics[width=\textwidth]{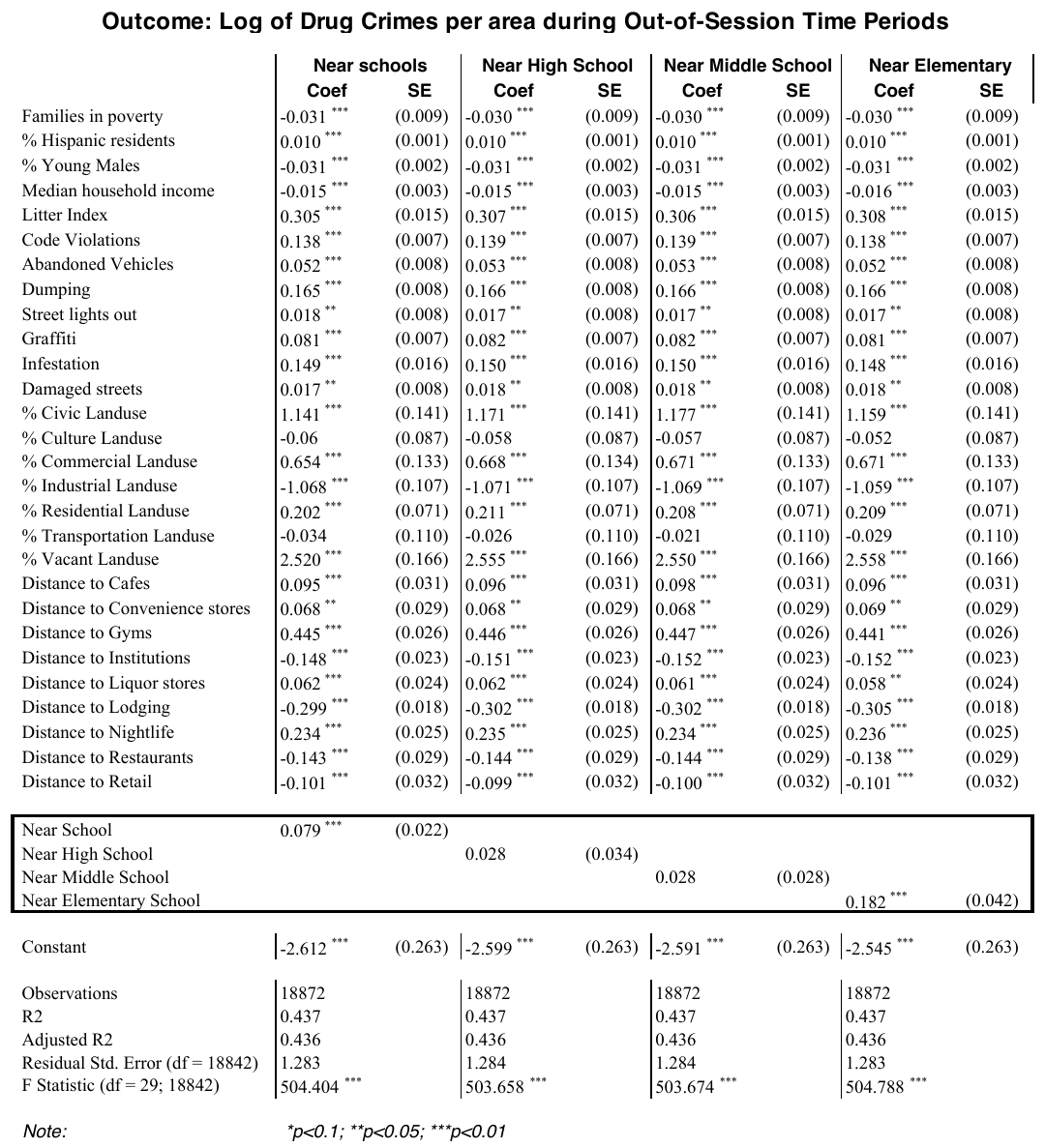}
\centering
\caption{Regression results with log of drug crimes per area during out-of-session time periods as outcome variable}
\label{reg-results-25-28-drug-outofsession}
\end{figure}

Figure \ref{reg-results-25-28-drug-outofsession} gives results for the regression models with log of drug crimes per area during out-of-session time periods as outcome variable.  

\begin{figure}[ht!]
\renewcommand\thefigure{S14}
\includegraphics[width=\textwidth]{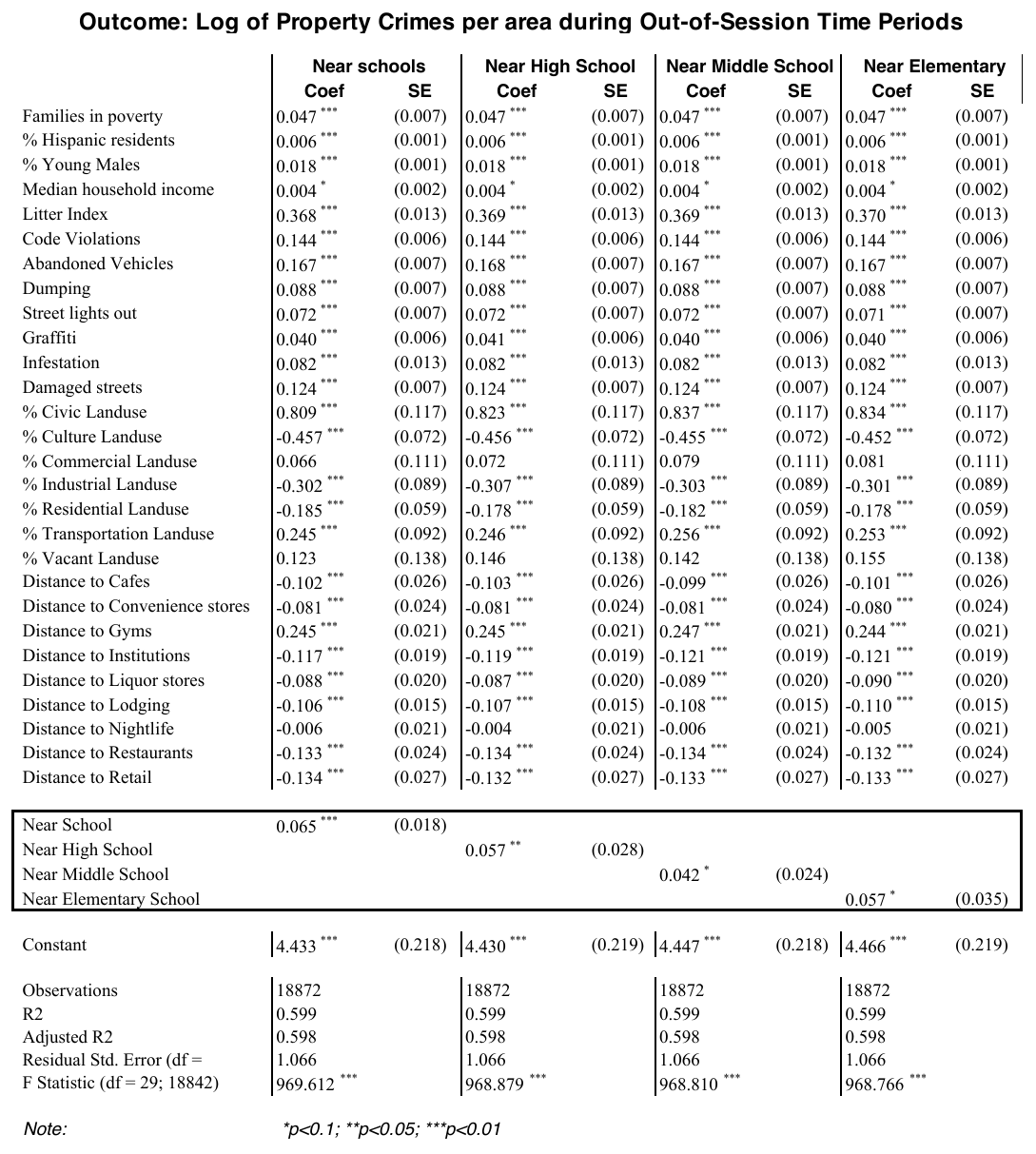}
\centering
\caption{Regression results with log of property crimes per area during out-of-session time periods as outcome variable}
\label{reg-results-29-32-property-outofsession}
\end{figure}

Figure \ref{reg-results-29-32-property-outofsession} gives results for the regression models with log of property crimes per area during out-of-session time periods as outcome variable.  

\newpage

\newpage

\newpage

\newpage

\newpage

\newpage

\newpage

\newpage

\end{document}